\documentclass[namedreferences]{SolarPhysics}
\usepackage[optionalrh,solaenum]{spr-sola-addons} % For Solar Physics 
\usepackage{graphicx}                    % For eps figures, newer & more powerfull
\usepackage{color}                       % For color text: \color command
\usepackage{url}                         % For breaking URLs easily trough lines
\newcommand{\aap}{    {\it Astron. Astrophys.}}
\newcommand{\aaps}{   {\it Astron. Astrophys. Suppl.}}

\newcommand{\apj}{    {\it Astrophys. J.}}

\newcommand{\pasp}{   {\it Publ. Astron. Soc. Pacific}}
\newcommand{\pasj}{   {\it Publ. Astron. Soc. Japan}}

\newcommand{\solphys}{{\it Solar Phys.}}

\begin{document}

\begin{article}

\begin{opening}

\title{Detection of Gravitational Redshift on the Solar Disk
by Using Iodine-Cell Technique\thanks{
Based on data collected at Hida Observatory, Kyoto University.\newline
Tables 3 and 4 are presented only as electronic materials.}}

%%%%%%%%%%%%%%%%%%%%%%%%%%%%%%%%%%%%%%%%%%%%%%%%%%%
%% Authors Names
%
\author{Y.~\surname{Takeda}$^{1}$ and 
        S.~\surname{Ueno}$^{2}$
       }

%%%%%%%%%%%%%%%%%%%%%%%%%%%%%%%%%%%%%%%%%%%%%%%%%%%
%% Runningheads
%
\runningauthor{Y. Takeda and S. Ueno}
\runningtitle{Detection of Solar Gravitational Redshift}

%%%%%%%%%%%%%%%%%%%%%%%%%%%%%%%%%%%%%%%%%%%%%%%%%%%
%% Affilations 
%
  \institute{$^{1}$ National Astronomical Observatory of Japan,
                    2-21-1 Mitaka, Tokyo 181-8588, Japan\\
                     email: \url{takeda.yoichi@nao.ac.jp}\\ 
             $^{2}$ Kwasan and Hida Observatories, Kyoto University,
                     Kurabashira, Kamitakara, Takayama, Gifu 506-1314, Japan\\
                     email: \url{ueno@kwasan.kyoto-u.ac.jp} \\
             }

%%%%%%%%%%%%%%%%%%%%%%%%%%%%%%%%%%%%%%%%%%%%%%%%%%%
%%% Abstract 
\begin{abstract}
With an aim to examine whether the predicted solar gravitational 
redshift can be observationally confirmed under the influence of the 
convective Doppler shift due to granular motions, we attempted measuring 
the absolute spectral line-shifts on a large number of points 
over the solar disk based on an extensive set of 5188--5212~$\rm\AA$ 
region spectra taken through an iodine-cell with the Solar 
Domeless Telescope at Hida Observatory. 
The resulting heliocentric line shifts at the meridian line
(where no rotational shift exists), which were derived by finding 
the best-fit parameterized model spectrum with the observed spectrum 
and corrected for the earth's motion, 
turned out to be weakly position-dependent as $\approx$+400~m~s$^{-1}$ 
near the disk center and increasing toward the limb up to 
$\approx$+600~m~s$^{-1}$ (both with a standard deviation 
of $\sigma \approx 100$~m~s$^{-1}$). 
Interestingly, this trend tends to disappear when the convective 
shift due to granular motions ($\approx -300$~m~s$^{-1}$ at the 
disk center and increasing toward the limb; simulated based on 
the two-component model along with the empirical center-to-limb
variation) is subtracted, finally resulting 
in the averaged shift of 698~m~s$^{-1}$ ($\sigma$ = 113~m~s$^{-1}$). 
Considering the ambiguities involved in the absolute wavelength 
calibration or in the correction due to convective Doppler shifts
(at least several tens m~s$^{-1}$, or more likely up to
$\lesssim 100$~m~s$^{-1}$), we may regard that this value is well 
consistent with the expected gravitational redshift of 633~m~s$^{-1}$.
\end{abstract}

%%%%%%%%%%%%%%%%%%%%%%%%%%%%%%%%%%%%%%%%%%%%%%%%%%%
%% Keywords
%

\end{opening}
%-------------------------------------------------

%%%%%%%%%%%%%%%%%%%%%%%%%%%%%%%%%%%%%%%%%%%%%%%%%%%
%% Sections
%
% \section{}%\label{s:?} 

%Section 1. Introduction
\section{Introduction}

Ever since Einstein (1911) predicted based on his general theory
of relativity that the wavelengths of solar spectrum lines should 
be shifted redward by a proportion of $2 \times 10^{-6}$ 
($\approx 600$~m~s$^{-1}$ in the velocity scale) as compared to the 
laboratory wavelengths, a number of solar physicists have tried to 
confirm this ``solar gravitational redshift'' observationally.
Although very few people nowadays doubt the practical validity of
this prediction from general relativity, given the remarkable
success in detecting this effect to a high precision from 
ground-based experiments, such as the well-known Pound--Rebka
experiment using the M\"{o}ssbauer effect (Pound and Rebka, 1960) 
or the experiment using two hydrogen masers on the ground and 
on a spacecraft (Vessot {\it et al.}, 1980), efforts of solar
astronomers toward confirming this gravitational redshift in the 
spectrum of the Sun have been non-terminating and continuing
up to now. Apart from the historical studies in the early time 
of first 50 years which are reviewed in Forbes (1961), several 
notable trials using modern techniques and/or observational data 
of higher quality have been done over the last 50 years,
as briefly summarized in Table 1.

While the authors of these studies in Table 1 reported
that they somehow detected the predicted redshift
even to a quantitatively satisfactory degree, their results 
are not yet necessarily convincing, since they do not seem to 
have properly taken into consideration the effect of convective 
shifts due to granular motions and the closely related problem of 
line-shift determinations ({\it i.e.}, definition of centroid wavelength) 
in case of asymmetric line profiles.
Besides, most of these investigations are based on observations
only on the disk center (or several disk points at best). 
Given that the well-known center-to-limb variation of 
the spectral line shift ({\it i.e.}, the shift increasing redward 
toward the limb, often called ``solar limb effect,'' first
discovered by Halm (1907)) is
regarded to be related with the photospheric velocity fields,
it is highly desirable to study absolute line-shifts 
on a {\it large} number of disk points while considering the 
convective shifts (varying from point to point) at the same time.
Namely, if the resulting residual shifts obtained from many points 
are confirmed to be nearly position-independent, it may suggest 
that shifts due to granular motions are successfully removed.

Recently, we carried out an extensive spectroscopic study
of solar differential rotation based on the spectra collected
at a large number ($\approx 3000$) of points over the solar disk 
while applying the iodine-cell technique for Doppler-shift measurements
(Takeda and Ueno, 2011; hereinafter referred to as Paper I). 
Since these data are suitable for the purpose mentioned 
above, we decided to conduct absolute Doppler shift measurements
by using these spectra, while paying special attention to their subset
observed on the solar meridian where the rotational shift is irrelevant,
in order to examine how and whether the solar gravitational redshift 
may be successfully confirmed by eliminating the shifts of 
velocity-field origin. This is the aim of this investigation. 

The remainder of this paper is organized as follows.
The observational data are explained in Section 2, and Section 3 
describes our adopted method of analysis for absolute line-shift
measurements. The resulting solutions and the simulation 
for the convective shifts to be subtracted are described in Sections 
4 and 5, respectively, based on which the final results for the 
solar gravitational redshift are discussed in Section 6.
The conclusion is summarized in Section 7.  
Finally, a supplementary section is presented as an appendix,
where the limb effect empirically studied by Balthasar (1984)
is reformulated for our application.

%Section 2. Observational Data (psfcalc)
\section{Observational Data}

We use the same observational data as adopted in Paper I, 
which were obtained on 20 and 21 July 2010 by using the 60~cm Domeless 
Solar Telescope (DST) with the horizontal spectrograph of the Hida 
Observatory of Kyoto University (Nakai and Hattori, 1985) and the 
iodine cell placed in front of the spectrograph slit. 
One sequence is made up of 13 scanning observations with a step of 
$\Delta r = R/12$ from the limb to the disk center at a fixed direction 
angle ($\theta$), and this sequence was repeated at 48 $\theta$-values 
(with an increment of $7.^{\circ}5$), resulting in 13$\times$48 points
on the disk as a total. The whole set was similarly conducted on each 
of the two days but with different slit setting; {\it i.e.}, E--W direction 
(20 July) and N--S direction (21 July). Since three spectra (at different
positions along the direction of the slit) were extracted from one 
observation, 3744 $(=13\times48\times3\times2)$ spectra in the 
5188--5212~$\rm\AA$ region (with reference I$_{2}$ molecular lines 
imprinted) covering the entire solar disk were obtained.
See Section 2 of Paper I for more details on the observational data. 
Prior to the analysis, a tentative wavelength calibration was done 
by using the representative solar lines, and an appropriate 
$\lambda$ (wavelength\footnote{Unless otherwise explicitly stated, 
we mean ``air wavelength in the standard condition'' by this term.}) 
vs. $x$ (pixel number) relation was assigned 
to each spectrum frame.

We examined the instrumental profile ($P_{\lambda}$) by fitting 
the observed lamp+I$_{2}$ spectrum ($D_{\lambda}^{\rm obs}$) 
with the modeled spectrum 
($D_{\lambda}^{\rm sim} \equiv T_{\lambda} \otimes P_{\lambda}$)
where $T_{\lambda}$ is the transmission function of I$_{2}$ vapor
and $\otimes$ means convolution.
Regarding $T_{\lambda}$, we adopt the spectrum (S/N~$\approx 1000$ with 
the resolving power of $R\approx 400000$) recorded for the I$_{2}$ cell 
at Lick Observatory in the 5000--6300~$\rm\AA$ region by using 
the Fourier Transform Spectrometer at the McMath telescope of 
the National Solar Observatories at Kitt Peak (Marcy and Butler, 1992), 
which was kindly provided by Dr. G. W. Marcy 
(see also Takeda {\it et al.}, 2002).

The instrumental profile was modeled by the sum of nine multiple Gaussian 
functions as $P(v) \propto \sum_{i=-4}^{4}{a_{i} \exp [(v - i)^2]}$
where $v$ (in km~s$^{-1}$) is measured from the line center. 
Assuming $a_{0} =1$, we determined the 8 coefficients ($a_{-4}$, 
$a_{-3}$, $a_{-2}$, $a_{-1}$, $a_{1}$, $a_{2}$, $a_{3}$, $a_{4}$)
by requiring the best fit between the modeled and observed spectra
in the 5202--5208~$\rm\AA$ region (Figure 1a). 
For a reference purpose, we also tried fitting with the Gaussian 
instrumental profile ($\propto \exp[-(v/v_{\rm m})^{2}]$) by finding 
the best-match solution of the parameter $v_{\rm m}$.
The shape of the resulting $P(v)$ is depicted in Figure 1b,
where we can see that the instrumental profile corresponding to our 
spectra (somewhat more rounded than the Gaussian) does not show any 
significant asymmetry that may affect absolute determinations of 
line shifts. The value of FWHM ($\approx 5$~km~s$^{-1}$) also suggests
that the wavelength resolution is $R \approx 60000$.

%Section 3. Method of Absolute Line-Shift Determination
\section{Method of Absolute Line-Shift Determination} 

%subsection 3.1 Spectrum Modeling
\subsection{Spectrum Modeling}

The iodine-cell technique has been extensively applied
for precise Doppler shift determinations so far ({\it e.g.}, 
spectroscopic search for exoplanets) since the pioneering study
of Marcy and Butler (1992) who showed how to accomplish high precision
Doppler analysis based on a stellar spectrum with many superimposed 
I$_{2}$ lines. However, its application has been limited essentially 
to ``relative'' measurements ({\it i.e.}, with respect to a template 
spectrum whose absolute wavelength scale does not need to 
be precisely calibrated) so far.
Since the determination of ``absolute'' shift is under question 
in this study, we have to modify Marcy and Butler's (1992) procedure 
(see Section 4 of their paper) appropriately. Namely, theoretically 
simulated spectrum with firmly established absolute wavelength scale 
must be used as the stellar template spectrum to be referenced.

Generally, in analogy with the profile modeling conventionally  
used in stellar physics, the theoretical solar intensity spectrum 
with imprinted I$_{2}$ lines at a disk point of direction cosine 
$\mu [\equiv \sqrt{1-(x^{2}+y^{2})/R^{2}}]$ may be expressed as follows:
\begin{equation}
I_{\rm th}(\lambda; \mu) \equiv 
k \{[I_{\rm intr}(\lambda; \mu; 
\xi,A_{1},A_{2},A_{3},\ldots) 
\otimes M_{\rm mt}(\lambda; \zeta)] \cdot
T_{{\rm I}_{2}}(\lambda)\} 
\otimes P(\lambda),
\end{equation}
where $k$ is an arbitrary normalization factor, $I_{\rm intr}(\lambda; \mu)$ 
is the intrinsic intensity at $\mu$ computed from a solar 
atmospheric model for given $\xi$ (microturbulence) and 
$A_{1},A_{2},A_{3},\ldots$ (elemental abundances), 
$M_{\rm mt}(\lambda; \mu; \zeta)$ is the broadening function 
corresponding to macroturbulence $\zeta$, $T_{{\rm I}_{2}}(\lambda)$ is 
the reference I$_{2}$ spectrum mentioned in Section 2, $P(\lambda)$
is the instrumental profile (Figure 1b), and ``$\otimes$'' denotes
convolution (while ``$\cdot$'' is usual multiplication).
Note that such a line-broadening formulation based on the dichotomous 
micro-/macro-turbulence is nothing but a too simple modeling of
actual velocity fields. Given that information for an adequate 
functional form of $M_{\rm mt}$ is lacking, we move $M_{\rm mt}(\lambda)$
to the outside of the square bracket in Equation (1) and further 
approximate $M_{\rm mt}(\lambda; \mu; \zeta) \otimes P(\lambda)$ by 
a single Gaussian macrobroadening function $G(\lambda; v_{\rm M}$),
which is parameterized by the macrobroadening parameter $v_{\rm M}$ 
($e$-folding width) as $G(v) \propto \exp(-v^{2}/v_{\rm M}^{2})$. 

Thus, we adopt the following model for line-spectrum simulation. 
\begin{eqnarray}
\lefteqn{I_{\rm th}(\mu, \lambda; k, \Delta\lambda_{\rm s}, \Delta\lambda_{{\rm I}_{2}}, v_{\rm M}, A_{1},A_{2},A_{3},\ldots)}\hspace{0.5cm}\\
 &\equiv&  k \{[I_{\rm intr}(\lambda+\Delta\lambda_{\rm s}; \mu; 
\xi,A_{1},A_{2},A_{3},\ldots) \cdot
T_{{\rm I}_{2}}(\lambda + \Delta\lambda_{{\rm I}_{2}})\} 
\otimes G(\lambda; v_{\rm M}). \nonumber
\end{eqnarray}
where $\Delta\lambda_{\rm s} (\equiv \lambda_{\rm s}^{\rm app} - \lambda_{\rm s}^{\rm ref})$ and $\Delta\lambda_{{\rm I}_{2}} (\equiv \lambda_{{\rm I}_{2}}^{\rm app} - \lambda_{{\rm I}_{2}}^{\rm ref})$ are the apparent wavelength shifts of solar lines 
and I$_{2}$ lines relative to the reference wavelength system ({\it i.e.}, 
wavelengths given Table 2 for the former or in Table 3 for the latter), 
respectively.

The parameters characterizing $I_{\rm th}$
($k, \Delta\lambda_{\rm s}, \Delta\lambda_{{\rm I}_{2}}, v_{\rm M}, A_{1},A_{2},A_{3},\ldots$) are determined by requiring that 
$\chi^{2} (\equiv \int [I_{\rm th}(\mu,\lambda) - I_{\rm obs}(\mu, \lambda)]^{2}d\lambda$ is minimized ({\it cf.} Takeda (1995) for detailed descriptions of 
the numerical procedure).
When the solutions of $\Delta\lambda_{\rm s}$ and $\Delta\lambda_{{\rm I}_{2}}$
have been established, the absolute line shift is obtained as
\begin{equation}
\Delta\lambda_{\rm abs} \equiv \Delta\lambda_{\rm s} - \Delta\lambda_{{\rm I}_{2}},
\end{equation}
since $\Delta\lambda_{{\rm I}_{2}}$ is the error (due to instrument instability, 
{\it etc.}) to be subtracted from $\Delta\lambda_{\rm s}$.

%subsection 3.2 Atomic Line Data for Spectrum Synthesis (specfit)
\subsection{Atomic Line Data for Spectrum Synthesis}

As mentioned in Section 3.1, spectrum synthesis calculation
for simulating the intrinsic specific intensity spectrum ($I_{\rm intr}$) 
is important, which needs to be well reproduce the observed solar 
spectrum by adequately adjusting the elemental abundances and the 
macrobroadening parameter. Regarding the calculation of 
$I_{\rm intr}(\lambda, \mu)$, we use Kurucz's (1993) WIDTH9 program, 
which was modified by Y. Takeda to enable spectrum synthesis 
by including many lines, along with Kurucz's (1993) ATLAS9 solar
photospheric model (LTE, plane-parallel model). In addition,
since the role of microturbulence is comparatively less significant 
in line broadening, $\xi = 1$~km~s$^{-1}$ is assumed as fixed.
As to the atomic line data (wavelengths, oscillator strengths, {\it etc.}) 
necessary for this simulation, we adopt the data compiled by Kurucz 
and Bell (1995). Since trial test calculations suggested 
the necessity of some adjustments, the following modifications 
were applied to in the $\log gf$ values of four lines.
Fe~{\sc i} 5197.929 ($\log gf = -1.64 \rightarrow -1.34$; 
increased by 0.3~dex), 
Fe~{\sc i} 5203.691 ($\log gf = -3.093 \rightarrow -9.99$; 
neglected), 
Ni~{\sc i} 5203.788 ($\log gf = -1.163 \rightarrow -9.99$; 
neglected), and 
Fe~{\sc i} 5209.884 ($\log gf = -3.26 \rightarrow -3.46$; 
decreased by 0.2~dex). 
Besides, since the weak depression at $\lambda \approx 5202 \rm\AA$ could not
be reproduced ({\it i.e.}, relevant line is missing in Kurucz and Bell's
database), we neglected (masked) this local feature in $\chi^{2}$ 
evaluation.

The finally adopted data for the important lines of appreciable 
strengths are summarized in Table 2, and how the observed 
disk-center spectrum (taken from the FTS solar spectrum atlas; 
Neckel 1994, 1999)\footnote{This spectrum atlas is available 
at $\langle$ ftp.hs.uni-hamburg.de/pub/outgoing/FTS-atlas $\rangle$.} 
can be fitted with the theoretical calculated $\mu=1$ intensity 
spectrum by using these atomic data (with appropriately 
adjusted elemental abundances and the broadening parameter) is 
displayed in Figure 2, where the lines given in Table 2 are also 
identified.

%subsection 3.3 Strengths and Formaton Depths of Spectral Lines
\subsection{Strengths and Formation Depths of Spectral Lines}

It is necessary to have knowledge on the strength and related quantities
({\it e.g.}, line-forming depth) of spectral lines existing in the 
relevant wavelength range (5188--5212~$\rm\AA$),
since the convective line shift (its correction is important 
as will be discussed in Section 5) appreciably depends on such
individual line properties. For this purpose, by using the solutions
of the elemental abundances derived as by-products of the synthetic
spectrum fitting in Section 3.2 (Figure 2), we {\it inversely}
computed EW (equivalent width at the disk center) for all lines
presented in Table 2, while assuming that each line is single
and isolated ({\it i.e.}, irrespective of whether they are actually
blended or not). Besides, we computed two kinds of line-formation
depth; one is the mean formation depth averaged over the profile
($\langle \log \tau \rangle$) and the other is the formation depth 
at the line-center ($\log \tau_{0}$), which are defined as follows.
\begin{equation}
\langle \log \tau \rangle 
\equiv \frac{\int R_{\lambda} \log \tau_{5000}(\tau_{\lambda} = 1) d\lambda} 
{\int R_{\lambda} d\lambda},
\end{equation}
where $R_{\lambda} (\equiv 1 - I_{\lambda}/I_{\rm cont})$ is the 
line depth of the emergent intensity profile, 
$\tau_{5000}(\tau_{\lambda} = 1)$ is the
continuum optical depth at 5000~$\rm\AA$ 
corresponding to $\tau_{\lambda} = 1$ at wavelength $\lambda$ 
within the profile, and the integration range is over the whole profile.
\begin{equation}
\log \tau_{0}  \equiv \log \tau_{5000}(\tau_{\lambda 0} = 1)
\end{equation}
where $\tau_{\lambda 0}$ is the optical depth at the line center.
(All these quantities are computed for the $\mu =1$ ray at the disk center.)
The resulting values of EW, $\langle \log \tau \rangle$, and 
$\log \tau_{0}$ are given in Table 2.

%subsection 3.4 Wavelength Scale of I_2 Reference Spectrum (i2spec_chk,nratio)
\subsection{Wavelength Scale of I$_{2}$ Reference Spectrum}

Regarding the reference spectrum of I$_{2}$ molecular lines ($T_{\lambda}$),
we adopt Marcy and Butler (1992)'s very high-resolution transmittance 
spectrum of Lick I$_{2}$ cell already mentioned Section 2.
This standard I$_{2}$ spectrum in the relevant wavelength region, 
which we also used to fit the spectrum of our I$_{2}$ cell (Figure 1a), 
is given in Table 3 (electronic data). Though the original data are 
represented in terms of the vacuum wavelength scale 
($\lambda_{\rm vac}$), we converted it into air wavelength 
($\lambda_{\rm air}$; which is usually used in optical spectroscopy
such as in Kurucz and Bell's atomic-line database we adopted) as
\begin{equation}
\lambda_{\rm air} = \lambda_{\rm vac}/n_{\rm air}   \; \; .
\end{equation}
Here, $n_{\rm air}$ is the refractive index of air in the standard 
condition represented as a function of the wavenumber $k$~cm$^{-1}$
[$= 10^{8}/\lambda_{\rm vac} (\rm\AA$)], for which we adopted the 
formula given in Cox (2000):
\begin{equation}
n_{\rm air} = 1. + 6432.8 \times 10^{-8} + 
2949810./(146.\times 10^{8}-k^{2}) + 
25540./(41.\times 10^{8}-k^{2}).
\end{equation}
While this Lick I$_{2}$ spectrum has so far been successfully used 
for very precisely detecting ``relative'' variations of radial velocity
intended for exoplanet search or stellar seismology, it is not clear
whether its ``absolute'' wavelength scale is sufficiently reliable.
Since this point is of primary importance in this study, we examined
its precision by comparing it with the theoretical I$_{2}$ spectrum 
by using ``IodineSpec'' program (Kn\"{o}ckel {\it et al.}, 2004), 
which can model the wavelengths of the rovibronic structure of the 
I$_{2}$ B--X spectrum with very high precision (its uncertainty 
is on the order of only $\approx 30$~MHz at $\approx 5200$~$\rm\AA$, 
which means a precision of $\approx 5\times 10^{-8}$ or 
$\approx 15$~m~s$^{-1}$; {\it cf.} their Figure 3), though the intensity 
distribution is not so precisely reproduced as wavelengths. 
The comparisons are displayed in Figures 3a (wide view) and 3b 
(magnified view of the 5196--5197~$\rm\AA$ region), where the vacuum 
wavelength scale ($\lambda_{\rm vac}$) is used in the abscissa. 
We notice from Figure 3b that the Lick I$_{2}$ spectrum appears to be 
slightly shifted redwards relative to the theoretical one. This is 
quantitatively confirmed by a cross-correlation analysis (Figure 3c), 
showing that the average shift in this 5196--5197~$\rm\AA$ bin amounts 
to 0.0028~$\rm\AA$ or $\approx 160$~m~s$^{-1}$. We similarly examined all 
twenty-four 1~$\rm\AA$-bins in the 5188--5212~$\rm\AA$ region
and the histogram of the resulting shifts are depicted in Figure 3d,
from which we conclude that the absolute wavelength scale of the
Lick I$_{2}$ spectrum is systematically redder and should be corrected 
blueward by $-167$~m~s$^{-1}$ (simple average of 24 values; with a 
standard deviation of 42~m~s$^{-1}$).

Besides, given that we transformed the vacuum wavelengths of reference 
I$_{2}$ spectrum into the air wavelengths by applying Equations (6) and 
(7), the air wavelengths of the atomic lines we adopted 
for simulating the solar spectrum have to be essentially on 
the same system. For the purpose of checking the consistency, we 
examined the $\lambda_{\rm vac}/\lambda_{\rm air}$ ratios (or the 
refractive index in the air; $n_{\rm KB}$) for all 719 lines 
(5188--5212~$\rm\AA$ region) included in Kurucz and Bell (1995) as
\begin{equation}
n_{\rm KB} \equiv 
(E_{\rm upper} - E_{\rm lower})/(10^{8} \lambda_{\rm air}),
\end{equation}
where $E_{\rm upper}$ and $E_{\rm lower}$ are the excitation potentials
of lower and upper levels in cm$^{-1}$, respectively, and 
$\lambda_{\rm air}$ is the air wavelength in $\rm\AA$ (all these 
three kinds of data are presented in Kurucz and Bell's database). 
The comparisons of such evaluated $n_{\rm KB}$'s and the corresponding
formula values ($n_{\rm formula}$) resulting from Equation (7) are 
depicted in Figure 4. We can recognize from this figure that 
the consistency is satisfactorily good and the average of the relative
difference [$\langle (n_{\rm KB} - n_{\rm formula})/n \rangle$] is 
only $-4.47\times 10^{-8}$ (with a standard deviation of 
$\sigma = 6.46\times 10^{-8}$) or $-13$~m~s$^{-1}$ 
($\sigma$ = 19~m~s$^{-1}$).\footnote{This dispersion ($\sigma$) of 
$\approx \pm 20$~m~s$^{-1}$ is reasonably attributed to 
the fact that the wavelengths of Kurucz and Bell's (1995) line data 
are presented only to the third decimal of $\rm\AA$ ({\it cf.} Table 2), 
which means that the intrinsic uncertainty amounting up to 
$\lesssim 0.0005$~$\rm\AA$ or $\lesssim 30$~m~s$^{-1}$ is inevitable 
in our analysis.} 
Thus, even though small, we formally apply a correction of 
$-(-13)$~m~s$^{-1}$ to the air wavelengths of I$_{2}$ lines 
({\it cf.} given 3rd column of Table 3) in order to maintain consistency 
with the wavelengths of atomic lines given in Table 2. 
(Regarding the sign of the correction, since 
$\lambda_{\rm air}^{\rm formula} < \lambda_{\rm air}^{\rm KB}$ 
because of $n_{\rm KB} < n_{\rm formula}$, we have to shift 
$\lambda_{\rm air}^{\rm formula}$ slightly redward to make it 
consistent with Kurucz and Bell's system.) 

Consequently, we conclude that the $\lambda_{{\rm I}_{2}}^{\rm ref}$ values 
given in Table 2, which we adopt as the reference (air) wavelengths 
of I$_{2}$ lines, should be corrected blueward by 
$\delta \lambda_{{\rm I}_{2}}^{\rm ref} = -154$~m~s$^{-1}$ 
(net combination of $-167$~m~s$^{-1}$ and $+13$~m~s$^{-1}$). 
More precisely, since the wavelengths of I$_{2}$ lines directly 
affect $\lambda_{{\rm I}_{2}}^{\rm ref}$ of $\Delta\lambda_{{\rm I}_{2}}  
(\equiv \lambda_{{\rm I}_{2}}^{\rm scale} - \lambda_{{\rm I}_{2}}^{\rm ref})$, 
it is included in $\Delta\lambda_{\rm abs}$ as the term of positive 
sign (+$\lambda_{{\rm I}_{2}}^{\rm ref}$; {\it cf.} Equation (3)), which means 
that  $\Delta\lambda_{\rm abs}$ should be corrected by 
+$\delta \lambda_{{\rm I}_{2}}^{\rm ref}$ = $-154$~m~s$^{-1}$.
In practice, the whole fitting analysis is conducted by literally 
using the (air) wavelengths of I$_{2}$ spectrum given Tables 3, 
and this correction is applied to the results at the last stage.

%Section 4. Results (fitsample,radvels,errorhisto)
\section{Results}

Although our primary interest is only on the meridian points 
($x \approx 0$) unaffected by the rotational velocity, we carried out 
our absolute wavelength-shift analysis for all the available data 
points over the entire solar disk, except for those at the limb 
because the solutions there were found to be comparatively uncertain 
and unstable ({\it i.e.}, the data at $r_{12}$ were not used; {\it cf.} Figure 2 
in Paper I).

Our spectrum covering 5188--5212~$\rm\AA$ was divided into four 
segments of 5--6~$\rm\AA$ [regions (a), (b), (c), and (d), 
{\it cf.} Table 2], and $\Delta\lambda_{\rm abs}^{i}$ was determined for 
each segment $i$ ($i$ = 1, 2, 3, 4; each corresponding to a, b, c, d, 
respectively) by applying the procedure
described in Section 3.1. An example of how the observed solar+I$_{2}$
spectrum matches the modeled spectrum corresponding to the best-fit 
solutions of the parameters is demonstrated in Figure 5. 
Then, the (raw) radial velocity of the observed point on the solar 
surface ($V_{\rm r}^{\rm raw}$) and its probable error ($\epsilon$) 
were computed as 
\begin{equation}
v_{i} \equiv c\Delta\lambda_{\rm abs}^{i}/\lambda_{i} -154 \; \;
({\rm m~s}^{-1}), 
\end{equation}
\begin{equation}
V_{\rm r}^{\rm raw} \equiv \frac{\sum_{i=1}^{N} v_{i}}{N}, 
\end{equation}
and
\begin{equation}
\epsilon \equiv \sqrt{\frac{\sum_{i=1}^{N}(v_{i} - V_{\rm r}^{\rm raw})^{2}}{N(N-1)}},
\end{equation}
where we applied the correction of $-154$~m~s$^{-1}$ due to the 
absolute wavelength scale problem of the adopted I$_{2}$ spectrum 
(Section 3.3), $c$ is the speed of light, and $N = 4$.
Further, we added the heliocentric correction for the earth's motion
($\Delta^{\rm hel}$; {\it cf.} Section 4.1 in Paper I) and obtained 
$V_{\rm r}^{\rm hel}$ in the heliocentric system as
\begin{equation}
V_{\rm r}^{\rm hel} = V_{\rm r}^{\rm raw} + \Delta^{\rm hel} .
\end{equation}
Figure 6 demonstrates, for the meridian case, how the results of 
$V_{\rm r}$ obtained at different observational times become mutually 
consistent by applying the $\Delta^{\rm hel}$ correction.

The distribution of the probable error ($\epsilon$) for each
measurement is shown in Figure 7, which is directly comparable
with Figure 9 of Paper I. As can be seen from Figure 7a,
$\epsilon$ mostly ranges between $\approx 50$~m~s$^{-1}$ and
$\approx 100$~m~s$^{-1}$; that is, by $\approx$~2--3 times larger
than the case of relative measurements in Paper~I, reflecting the 
difficulty of absolute measurements. We do not see any clear dependence
of $\epsilon$ upon the position ($r$) on the disk, except for a slight 
decrease near to the limb (Figure 7b).
The results of our analysis [observational time, ($r, \theta$) 
as well as ($x,y$)~coordinates, $V_{\rm r}^{\rm raw}$, $\Delta^{\rm hel}$, 
and $V_{\rm r}^{\rm hel}$] are presented in Table 4 (electronic data) 
for the meridian case of our concern, which will be discussed in Section 6.

%Section 5. Simulation of Convective Blue-Shift (sim2comp,simulation)
\section{Simulation of Convective Blue-Shift}

\subsection{Line Shift due to Convective Granular Motions}

Although we have derived the absolute line-shift ($V_{\rm r}^{\rm hel}$) 
reduced to the rest system of the Sun at each point on the disk,
it alone is still insufficient for our main purpose of quantitatively
confirming the gravitational redshift, since the systematic shift 
due to convective velocity fields has still to be subtracted.

A hot up-rising bright bubble (granular cell) originating from the 
deeper convective zone gradually decelerates its speed as it moves 
upward while dissipating its energy, and eventually stops and turns 
into cool downflows (dark intergranular lane). Since the contribution 
of the brighter former is larger than the darker latter in the 
integrated spectrum, the net effect generally makes a blue shift 
as seen from the observer. Since the extent of velocity contrast 
is depth-dependent ({\it i.e.}, becoming less significant toward higher 
layers), such a blue shift tends to be more appreciable for weak 
deep-forming lines than strong high-forming lines, as seen in 
Figure 2 of Allende Prieto and Garc\'{\i}a L\'{o}pez (1998). 
For the same reason, the off-center region (forming deeper) 
of a line profile is more blue-shifted than the region of line-center 
core (forming higher), which results in an asymmetric profile with 
a ``C''-shaped bisector (see, {\it e.g.}, Dravins, 1982).

To make things further more complicated, not only the vertical 
velocity field but also the horizontal field has to be considered 
when studying the variation of the convective shifts over the disk.
This results in a flatter decreasing rate of blue shift (from disk center
toward the limb) compared to the case of the vertical field only;
actually, even a peak of blue shift is occasionally observed not the 
disk center but at off-center around $\mu \approx 0.8$ (Schr\"{o}ter, 1959;
Beckers and Nelson, 1978; Balthasar, 1984, 1985). 

Considering this complexity, we decided to adopt the following strategy:
As far as the convective shift at the disk center ($\mu = 1$) is
concerned, we can evaluate it for any relevant line to a sufficient
precision, thanks to the availability of a well-tuned two-component model.
Meanwhile, since this kind of line shift at positions off the center
($0 < \mu < 1$) is too difficult to simulate, we invoke 
the empirical work of Balthasar (1984), who extensively investigated the 
relative $\mu$-dependence of the shifts for many spectral lines of different 
properties and classified them in terms of line-formation depth,
which we reformulated in a more easy-to-use manner ({\it cf.} the Appendix). 
Accordingly, as successively described in the following sections, 
we (1) first simulate the absolute line-shifts at the disk center 
($\mu =1$) for representative lines (Section 5.2), 
and (2) then combine them with the relative center-to-limb 
trends empirically elucidated by Balthasar (1984) (Section 5.3).

\subsection{Absolute Shift at the Disk Center}

For the purpose of simulating the disk-center intensity profile under 
convective velocity fields, we adopt the two-component model proposed 
by Borrero and Bellot Rubio (2002). Their model consists of two kinds 
of plane-parallel atmospheric models (with only vertical velocity
fields) corresponding to hot/bright granular (g) and cool/dark 
intergranular (i) region. The details
are presented in Appendix A in their paper, where $T$ (temperature), 
$P_{\rm e}$ (electron pressure), $P_{\rm g}$ (gas pressure), 
$\rho$ (density), and $v$ (vertical velocity) are given
as functions of depth (along with the depth-independent 
microturbulence and macroturbulence) for each of the two models.
The structures of $T(\tau_{5000})$ and $v(\tau_{5000})$ for each model
are depicted in Figure 8a. 

By using the WIDTH9 program (as in Section 3) along with these models, 
we computed the emergent specific intensity profile 
$I^{\rm g}(\lambda)$ and $I^{\rm i}(\lambda)$ at the disk center, 
from which the final spectrum is eventually obtained as the 
combination of these two:
\begin{equation}
I^{\rm comb}(\lambda) \equiv 
(1-\alpha)I^{\rm g}(\lambda) + \alpha I^{\rm i}(\lambda),
\end{equation}
where $\alpha$ is the filling factor assumed to be 0.24 according to
Figure 3 of Borrero and Bellot Rubio (2002).

In order to check our computational procedure, we simulated the 
$\mu = 1$ profiles of 22 Fe~{\sc i} lines given in Table 1 of 
Borrero and Bellot Rubio (2002), and confirmed that most\footnote{
More precisely, the predicted shifts of the 3 lines (Fe~{\sc i}~6574.2285, 
Fe~{\sc i}~6581.2100, and Fe~{\sc i}~6739.5220) out of 22 lines
appear to be only slightly shifted bluewards compared to the observation; 
otherwise, the profile/wavelength match for the remaining lines 
was found to be very good.} of the resulting 
line-shifts satisfactorily reproduced those of the observed line
profiles of Neckel's (1994, 1999) solar disk-center spectrum atlas.
Figure 8b shows an example of such simulated $I^{\rm g}$, $I^{\rm i}$, 
and $I^{\rm comb}$ (each being normalized with respect to the continuum 
level of its own) for the representative Fe~{\sc i} 5194.941 line.

Such computed  $I^{\rm comb}(\lambda)$ is further convolved with 
the instrumental profile ($P$; {\it cf.} Section 2), resulting in an 
asymmetric broadened profile corresponding to our instrumental 
condition. Profile examples before and after I.P. convolution are 
displayed in Figure 8c again for the case of Fe~{\sc i} 5194.942. 

The final step is to determine the line shift from such simulated
(asymmetric) profile by applying the same procedure described in 
Section 3; {\it i.e.}, fitting with the classically modeled symmetric
profile  while adjusting the line shift ($\delta \lambda$ or 
$\delta V$) along with the broadening width and the elemental 
abundance. This process is demonstrated in Figure 8d, where we 
can see that the match is satisfactory.

We applied the whole above-mentioned procedure to 18 lines of 
appreciable strengths, which we selected by the criterion of 
$\langle \log \tau \rangle < -1$. 
Such determined line shifts ($\delta V_{1}$) at the disk center
($\mu = 1$) for these lines are given in the last column of Table 2,
where we can see that $\delta V_{1}$ is in the range of
$-310$~m~s$^{-1} \lesssim \delta V_{1} \lesssim -230$~m~s$^{-1}$.\footnote{
It should be stressed that we adopted the fitting-based line-shift 
($\delta V_{\rm fit}$), while the line shift ($\delta V_{\rm min}$) 
defined in terms of the wavelength of minimum intensity has been more 
commonly used so far. We also determined $\delta V_{\rm min}$
by interpolating the lowest three profile points by a quadratic function
(see, {\it e.g.}, Allende Prieto and Garc\'{\i}a L\'{o}pez, 1998) for all 
these 18 lines in the present case, and found that the difference
$\delta V_{\rm min} - \delta V_{\rm fit}$  tends to be EW-depedent,
which varies from $-60$~m~s$^{-1}$ (EW~$\approx 70$~m$\rm\AA$) to 
$+120$~m~s$^{-1}$ (EW~$\approx 230$~m$\rm\AA$) and can be  
approximated by the linear-regression relation 
$\delta V_{\rm min} - \delta V_{\rm fit} \; ({\rm m\;s^{-1}}) $ = 
$- 129 + 1.02 {\rm EW} \; ({\rm m \AA})$. Since the typical extent of this 
difference is several tens of m~s$^{-1}$, we may state that the
results are not seriously influenced by the definition of line shift.}

\subsection{Center-to-Limb Variation of $\delta V$}

We are now ready to derive the $\mu$-dependent convective shift,
$\delta V_{\mu}$, by combining the disk-center $\delta V_{1}$
computed in Section 5.1 and the center-to-limb variation formulated
in the Appendix based on the results of Balthasar (1984). 

For a line $k$,  we first evaluate $\log \tau_{\rm B}^{k}$
(Balthasar's core-forming depth)
from $\log \tau_{0}^{k}$ (our line-center-forming depth) 
by solving Equation (23). Then, the coefficients $b^{k}$ and $c^{k}$ 
in the velocity unit can be derived from Equations (21),
which are sufficient with Equation (19) to establish the relative 
$\mu$-dependence of $\delta V$ as
\begin{equation}
\delta V^{k}(\mu) - \delta V^{k}(1) 
= \Delta v_{\rm B}^{k}(\mu) - \Delta v_{\rm B}^{k}(1) 
= b^{k} (1-\mu) + c^{k} (1-\mu)^2.
\end{equation}
Therefore, combining Equation (14) with $\delta V_{1}^{k}$ which was 
already derived in Section 5.2, we can express the absolute 
convective shift for line $k$ at any $\mu$ as
\begin{equation}
\delta V_{\mu}^{k} = \delta V_{1}^{k} + b^{k} (1-\mu) + c^{k} (1-\mu)^2.
\end{equation}
Such obtained position/angle-dependent $\delta V_{\mu}^{k}$ results
for the 18 lines are shown in Figure 11a ($\delta V$ vs. $\mu$) and 
11b ($\delta V$ vs. $\sqrt{1-\mu^{2}}$), where we can see that
the tendencies are nearly similar to each other, except for
two conspicuously strong Cr~{\sc i} lines (at 5206.038 and 5208.419~$\rm\AA$)
of low excitation ($\chi_{\rm low} \approx 1$~eV).

As the final step, we average these $\delta V_{\mu}$ results
over 18 lines in order to obtain the mean relation, while adopting 
\begin{equation}
w^{k} \propto 1/\langle \tau^{k} \rangle \equiv 
10^{-\langle \log \tau^{k} \rangle}   
\end{equation}
as the weight\footnote{
This choice of weight ($\propto 1/\langle \tau \rangle$) is 
for the following reason.
The precision of radial velocity determination is higher not only 
for stronger (deeper) but also for sharper (narrower) lines,
which means that the radial velocity is mainly determined
by sharp saturated lines at the flat part of the curve of growth.
Meanwhile, ``very'' strong lines with appreciable damping 
wings do not necessarily contribute so significantly because of
their less sharpness. We point out here that the extent of saturation 
({\it i.e.}, both of strength and sharpness), to which more weight should
be given, is most well described by $1/\langle \tau \rangle$
because it increases as a line gets stronger from the linear to the flat
part of the curve of growth, while it then turns to decrease as a line
further becomes very stronger and comes into the damping part of 
the curve of growth, which is the characteristic just what we want. 
(In contrast, $1/\tau_{0}$ is ever increasing monotonically
with the growth of line strength.) Figure 11d describes well
this situation.
} for line $k$.

Accordingly, we finally obtain 
\begin{eqnarray}
\langle \delta V_{\mu} \rangle 
&=& \langle \delta V_{1} \rangle + \langle b \rangle (1-\mu) 
+ \langle c \rangle (1-\mu)^2  \nonumber \\
&=& -278.5 + 79.9 (1-\mu) + 422.3 (1-\mu)^2 \;\; ({\rm m \; s}^{-1}),
\end{eqnarray}
where $\langle \cdots \rangle$ is the $w^{k}$-weighted average over 18 lines.
Such obtained $\langle \delta V_{\mu} \rangle$, which is also
depicted in Figures 11a and 11b by the thick red line, will be used 
in Section 6 for correcting $V_{\rm r}^{\rm hel}$ for the convective shift.

%Section 6. Discussion (all_comb)
\section{Discussion}

We now discuss the results of absolute line shifts obtained in Section 4, 
while taking into account the convective-shift correction estimated 
in Section 5. In the following description, $x$ and $y$ denote the 
Cartesian coordinates on the solar disk ({\it cf.} Figure 2a in Paper I).
The relations between $V_{\rm r}^{\rm hel}$ obtained for all 
the observed points and each of the positional coordinates
are displayed in Figures 10a ($V_{\rm r}^{\rm hel}$ vs. $y$)
and 10c ($V_{\rm r}^{\rm hel}$ vs. $x$), where the exhibited trends 
can be reasonably understood by the Doppler shift due to solar rotation 
($V_{\rm r}^{\rm hel} \propto x$; {\it cf.} Section 4.1 in Paper I). 
A rough inspection of these two figures suffices to realize that
$V_{\rm r}^{\rm hel}$ distributes around $\approx +500$~m~s$^{-1}$
at $x \approx 0$ (meridian line) or near to the disk center (where 
the data points are densely confined), which suggests that 
$V_{\rm r}^{\rm hel}$ approaches the expected 
gravitational shift\footnote{This value of 633~m~s$^{-1}$ corresponds 
to an observer on the earth while the redshift is 636~m~s$^{-1}$
for the observer at infinity.} of $+633$~m~s$^{-1}$ where 
the rotational Doppler shift is negligible.

In order to study this situation more in detail, we pay attention
to the selected dataset corresponding to the observations
on the meridian line ($x \approx 0$), which are given in Table 4.
The $V_{\rm r}^{\rm hel}$ vs. $y$ correlation for this subset
is shown in Figure 10b, where we can confirm that $V_{\rm r}^{\rm hel}$ 
values confine around $\approx +500$~m~s$^{-1}$ (with a standard 
deviation of $\sigma \approx 100$~m~s$^{-1}$) as mentioned above. 
In addition, we can recognize a tendency of slight increase 
of $V_{\rm r}^{\rm hel}$ with $|y|$; {\it i.e.}, $\approx$+400~m~s$^{-1}$ 
at the disk center and increasing toward the limb up to 
$\approx$+600~m~s$^{-1}$. (Actually, such a change of
$V_{\rm r}^{\rm hel}$ from the limb to the center on the meridian 
line is already observed in Figure 6.) 
This trend is nothing but the ``limb effect'' already mentioned
in (see, {\it e.g.}, Figure 2 of Dravins, 1982).
Interestingly, when we overplot the $\langle \delta V \rangle$ vs. $y$ 
relation according to Equation (17) [describing $\langle \delta V\rangle $ 
as a function of $\mu$ ($= \sqrt{1-(y/R)^{2}}$] derived in Section 5, 
its tendency is found to be similar to the behavior of $V_{\rm r}^{\rm hel}$
({\it cf.} the solid line in Figure 10b). Thus, $V_{\rm r}^{\rm hel}$ is
corrected for $\langle \delta V \rangle $ as
\begin{equation}
V_{\rm r}^{\rm hel,corr} \equiv V_{\rm r}^{\rm hel} - \langle \delta V \rangle,
\end{equation}
and the resulting $V_{\rm r}^{\rm hel,corr}$ is plotted against $y$
in Figure 10d. We can see in this figure that the slight curvature
observed in Figure 10b is more or less removed by this correction
(especially in the northern hemisphere, though somewhat overcorrected
in the southern hemisphere)
and $V_{\rm r}^{\rm hel,corr}$ is now nearly $y$-independent
along the meridian line, which is just what we intended to 
achieve ({\it i.e.}, to consistently detect the gravitational redshift
at a number of different points by eliminating the position-dependent 
limb effect; {\it cf.} Section 1). 
The distribution histogram of $V_{\rm r}^{\rm hel,corr}$ and the 
$\langle V_{\rm r}^{\rm hel,corr}\rangle$ (locally averaged value) 
vs. $y$ relation are shown in Figures 10e and 10f, respectively.\footnote{
It is interesting to note that our  
$\langle V_{\rm r}^{\rm hel,corr}\rangle$ of 696~m~s$^{-1}$ 
at the disk center ({\it cf.} Figure 10f) is comparable to 684~m~s$^{-1}$ 
derived by Balthasar (1985) as the asymptotic value for the absolute 
disk-center line-shift at the strong-line limit of 
$\log\tau_{\rm B} \rightarrow -\infty$
(where the convective shift is expected to be insignificant).} 
The average of all the $V_{\rm r}^{\rm hel,corr}$ (meridian) data
turned out to be 
$\langle\langle V_{\rm r}^{\rm hel,corr}\rangle\rangle = 698$~m~s$^{-1}$ 
(with a standard deviation of $\sigma = 113$~m~s$^{-1}$). 

Here, it may worthwhile to recall that several sources of errors 
are involved in our results: 
Each individual absolute $V_{\rm r}$ measurement is accompanied 
by a probable error of $\approx 50$--100~m~s$^{-1}$ (Figure 7). 
Also, the randomness of granular velocity fields (represented by
``wiggly-line'' character seen in high-resolution slit-spectrograms) 
produces an appreciable dispersion of $V_{\rm r}$ (even though
it must have been somewhat smoothed because spatial information over 
42$''$ has been averaged in our spectrum; {\it cf.} Section 2 in Paper I),
which is estimated in the present case to be on the order of 
$\approx 100$~m~s$^{-1}$ (Figures 10e and 10f). Yet, since these
are random errors, they could (in principle) be reduced in the 
averaged $\langle\langle V_{\rm r}\rangle\rangle$ by summing up as many 
data as possible; {\it e.g.}, averaging of all 144 data on the meridian 
would result in a reduction by a factor of $\approx$~10 ($\approx \sqrt{144}$).
We thus estimate that the error component in 
$\langle\langle V_{\rm r}\rangle\rangle$
due to such random sources would have been reduced to the 
$\approx$~10--20~m~s$^{-1}$ level.
Actually, however, systematic errors must be more important, which are
hard to evaluate. For example, while we corrected the absolute 
wavelength-scale of Lick I$_{2}$ spectrum by 162~m~s$^{-1}$, this 
value itself already has an uncertainty with a magnitude of several 
tens m~s$^{-1}$ (Figure 3d). Besides, although our convective-shift 
corrections turned out seemingly successful, we have no idea about
how much systematic errors are involved in these, given that 
this correction differs from line to line with variations of
$\approx$~50--100~m~s$^{-1}$ ({\it cf.} Figure 9).
Considering all what described above, we would modestly state that
our final $\langle\langle V_{\rm r}^{\rm hel,corr}\rangle\rangle$ 
value suffers inevitable ambiguities of no less than several tens 
m~s$^{-1}$ or maybe amounting up to $\lesssim 100$~m~s$^{-1}$.

Accordingly, while the resulting 
$\langle\langle V_{\rm r}^{\rm hel,corr}\rangle\rangle$ 
of 698~m~s$^{-1}$ is slightly larger than the predicted gravitational 
redshift of 633~m~s$^{-1}$ by +65~m~s$^{-1}$, this difference 
is not meaningful in view of the expected uncertainties of 
several tens m~s$^{-1}$ to $\lesssim 100$~m~s$^{-1}$ as estimated above. 
We thus conclude that the solar gravitational redshift has been 
quantitatively confirmed in a reasonable manner by our absolute 
line-shift analysis using the iodine-cell technique, in combination 
with the simulated convective shift corrections.

%Section 7. Conclusion
\section{Conclusion}

Motivated by the situation that various past studies of solar 
gravitational redshift are not convincing because few of them
seem to have properly considered the Doppler shifts due to 
inhomogeneous convective motions,
we decided to examine whether the predicted solar gravitational 
redshift can be observationally confirmed by removing the 
appropriately simulated convective shift.

For this purpose, we attempted measuring the absolute spectral 
line-shifts on a large number of points over the solar disk 
based on an extensive set of 5188--5212~$\rm\AA$ region spectra 
taken through an iodine-cell with the Solar Domeless Telescope 
at Hida Observatory, where we paid attention only to the selected
dataset observed on the solar meridian where the rotational Doppler
shift is irrelevant.

In analogy with the analysis technique for precise ``relative'' 
shift measurements (from I$_{2}$+object spectra) elaborated
by Marcy and Butler (1992), we developed a method for measuring
``absolute'' shifts by using the theoretical solar intensity spectrum
template, which is computed from a model atmosphere and parameterized
by elemental abundances along with the broadening parameter; 
and the absolute shift can be established by finding the best-fit 
solution of the parameterized model spectrum with the observed spectrum.

The resulting heliocentric absolute shifts ($V_{\rm r}^{\rm hel}$)
on the meridian line turned out to be weakly position-dependent 
as $\approx$+400~m~s$^{-1}$ near the disk center and increasing toward 
the limb up to $\approx$+600~m~s$^{-1}$ (both with a standard deviation
of $\approx 100$~m~s$^{-1}$). 

We then simulated the convective blue shifts for representative 18 lines 
by using Borrero and Bellot Rubio's (2002) two-component model 
for the disk center, combined them with the empirical relations of
relative center-to-limb variation investigated by Balthasar (1984),
and finally adopted the weighted-average over 18 lines.  
The resulting shift is $\langle \delta V \rangle \approx -300$~m~s$^{-1}$ 
at the disk center, gradually increases to $\approx -100$~m~s$^{-1}$ at 
$|y/R| \approx 0.9$, and shows a suddenly increase up to $\approx +200$~m~s$^{-1}$ 
at the limb ($|y/R| = 1$).

Interestingly, the position-dependence of $V_{\rm r}^{\rm hel}$ tends to 
be removed when the simulated convective blue-shift is subtracted, which 
finally makes the averaged shift for all the meridian data to be 
$\langle\langle V_{\rm r}^{\rm hel,corr}\rangle\rangle$ = 698~m~s$^{-1}$ 
($\sigma$ = 113~m~s$^{-1}$). 

Considering the ambiguities involved in the absolute wavelength 
calibration or in the correction due to convective Doppler shifts
(at least several tens m~s$^{-1}$, or more likely up to
$\lesssim 100$~m~s$^{-1}$), we may state that this value is well 
consistent with the expected value of +633~m~s$^{-1}$ (due to
the gravitational potential on the solar surface) as predicted 
from the general relativity theory. This means that we could 
confirm the solar gravitational redshift in a quantitatively 
satisfactory level within the precision of our analysis.

\begin{acks}
Y. Takeda heartily thanks Dr. H. Kn\"{o}ckel for kindly providing
the ``IodineSpec'' program, which turned out very helpful
for checking the absolute wavelength scale of the reference 
I$_{2}$ spectrum. 
\end{acks}

\appendix
\section*{Analytical Formulation of Balthasar's (1984) Empirical 
Limb-Effect}

Balthasar (1984) carried out an empirical investigation on
the center-to-limb variation of spectral-line shift on the solar disk
and found that the trend of line-asymmetry as well as of line-shift 
depends on the formation heights of line cores ($\log \tau_{\rm B}$). 
He tried to fit the observed shifts of many spectral lines
by a quadratic polynomial in terms of $1- \mu$ 
 ($\mu \equiv \cos \theta$: direction cosine) as 
\begin{equation}
\Delta\lambda_{\rm B} \; ({\rm or}\; \Delta v_{\rm B}) 
= a + b (1- \mu) + c (1 - \mu)^{2}.
\end{equation}
and determined the coefficients ($a, b, c$) for 6 different 
$\log \tau_{\rm B}$ ranges ({\it cf.} Table IV therein),
as shown in Figures 11a--11c. 
For practical convenience, we derived the analytical expression of 
$a$, $b$, and $c$ with respect to $\log \tau_{\rm B}$ by applying 
quadratic least-square fitting to these data as
\begin{eqnarray}
a \; ({\rm m\AA}) &=& 
0.254 + 0.619 (\log \tau_{\rm B}) + 0.138 (\log \tau_{\rm B})^{2}, \nonumber \\
b \; ({\rm m\AA}) &=& 
-7.03 + 0.00450 (\log \tau_{\rm B}) + 0.819 (\log \tau_{\rm B})^{2}, \\
c \; ({\rm m\AA}) &=& 
18.4 + 1.90 (\log \tau_{\rm B}) - 0.530 (\log \tau_{\rm B})^{2}, \nonumber
\end{eqnarray}
as depicted in these figures.
Considering that Balthasar (1984) used 5000~$\rm\AA$ as the standard
wavelength, we can convert these coefficients into velocity scale as
\begin{eqnarray}
a \; ({\rm m \, s^{-1}}) &=& 
15.3 + 37.1 (\log \tau_{\rm B}) + 8.30 (\log \tau_{\rm B})^{2}, \nonumber \\
b \; ({\rm m \, s^{-1}}) &=& 
-421.8 + 2.70 (\log \tau_{\rm B}) + 49.1 (\log \tau_{\rm B})^{2}, \\
c \; ({\rm m \, s^{-1}}) &=& 
1100. + 113.9 (\log \tau_{\rm B}) - 31.8 (\log \tau_{\rm B})^{2}. \nonumber 
\end{eqnarray}

One problem is that Balthasar (1984) did not clarify the definition 
of his ``line-core formation depth'' (expressed as ``lg$\tau$'' therein
but denoted here as $\log \tau_{\rm B}$).
We computed $\langle \log \tau \rangle$ and $\log \tau_{0}$ 
(the mean formation depth and the line-center formation depth 
defined in Equations (4) and (5)) for all 143 lines given in Table~I 
of Balthasar (1984) in the same way as we did in Section 3.3. 
These results are compared with his $\log \tau_{\rm B}$ values
in Figures 11d and 11e, where we can see that his $\log \tau_{\rm B}$
is similar to our $\log \tau_{0}$, though both are not in a
linear relationship with each other.
We found from least-squares analyses the following relations:
\begin{eqnarray}
\langle \log \tau \rangle &=& 
-0.0375 + 1.03 (\log \tau_{\rm B}) + 0.124 (\log \tau_{\rm B})^{2}, \\
\log \tau_{0} &=& 
 0.0136 + 1.54 (\log \tau_{\rm B}) + 0.111 (\log \tau_{\rm B})^{2}.
\end{eqnarray}
In the practical application in Section 5, our $\log \tau_{0}$ is 
converted into $\log \tau_{\rm B}$ by using Equation (23)
for each line, from which the necessary coefficients (such as 
$b$ or $c$) can be derived by applying Equation (21).

%Table 1
%\clearpage
\setcounter{table}{0}
\begin{table}[h]
\scriptsize
\caption{Solar gravitational redshift determinations over the last 
50 years (expressed by $R \equiv \Delta_{\rm obs}/\Delta_{\rm the}$).}
%\begin{center}
\begin{tabular}
{l}\hline 
Authors: Blamont and Roddier (1961) \\
Method: Sr~{\sc i}~4607 line with atomic-beam resonant scattering method \\
Results: $R \approx 0.5$ ($\approx$~5~m$\rm\AA$/9.76~m$\rm\AA$) [disk center],
$R \approx 1.0$ ($\approx$~10~m$\rm\AA$/9.76~m$\rm\AA$) [limb] \\
Remark: Redshift of 2.4 m$\rm\AA$ due to Stark effect is subtracted \\
\hline
Authors: Snider (1970) \\
Method: K~{\sc i}~7699 line with atomic-beam resonant scattering method  \\
Results: $R = 0.61\pm 0.06$ ($10 \pm 1$~m$\rm\AA$/16.3~m$\rm\AA$) [disk center] \\
\hline
Authors: Snider (1972) \\
Method: K~{\sc i}~7699 line with atomic-beam resonant scattering method \\
Results: $R = 1.01\pm 0.06$ ($16.4 \pm 1$~m$\rm\AA$/16.3~m$\rm\AA$) [disk center] \\
\hline
Authors: Beckers (1977) \\
Method: Measuring line-shifts of Zeeman-insensitive ($g=0$) Ti~{\sc i} 5713 
line in the umbrae of 7 sunspots \\
Results: $R = 0.97 \pm 0.02$ ($613 \pm 14$~m~s$^{-1}$/633~m~s$^{-1}$) \\
\hline
Authors: LoPresto {\it et al.} (1980) \\
Method: 738 Fe lines based on Pierce and Breckinridge's (1972) 
wavelength table (disk center) \\
Results: $R = 0.76 \pm 0.24$ (all 738 lines at 2930--5762~$\rm\AA$),
$R = 0.97 \pm 0.16$ (only 74 strong lines at 3075--4958~$\rm\AA$) \\
\hline
Authors: LoPresto {\it et al.} (1991) \\
Method: Wavelength shifts (at the profile core)
of oxygen triplet lines at 7772--7775~$\rm\AA$ \\
Results: $R = 0.73$ (460~m~s$^{-1}$/633~m~s$^{-1}$) [disk center],
$R = 0.99$ (627~m~s$^{-1}$/633~m~s$^{-1}$) [limb] \\
Remark: Measurements on 7 different ƒÊ points
suggest a tendency of increasing shifts toward the limb \\
\hline
Authors: Cacciani {\it et al.} (2006) \\
Method: Doppler shift measurements over a number of points over the disk
by using a magneto-optical filter \\
Results: $R = 0.99$ (625~m~s$^{-1}$/633~m~s$^{-1}$) [at the meridian line of zero-rotational velocity] \\
Remark: Several assumptions are involved regarding other sources causing line shifts
({\it cf.} section 6.1 therein) \\
\hline
\end{tabular}
%\end{center}
Note: The theoretical solar gravitational redshift of 633~m~s$^{-1}$ 
(the value corresponding to observations from the earth)
is adopted here.
\end{table}

%Table 2
%\clearpage
\setcounter{table}{1}
\begin{table}[h]
\tiny
%\scriptsize
%\small
\caption{Atomic and relevant data of important lines at each region of spectrum fitting.}
\begin{center}
\begin{tabular}
{cccrcccrccl}\hline 
Species & $\lambda_{\rm air}$  & $\chi_{\rm low}$  & $\log gf$ & Gammar & Gammas & Gammaw &
EW  & $\langle\log \tau\rangle$ & $\log \tau_{0}$ & $\delta V_{1}$\\
 & ($\rm\AA$) & (eV) &  &  &  &  & (m$\rm\AA$) &  &  & (m~s$^{-1}$)\\
\hline
\multicolumn{10}{l}{Region (a): 5188.5--5193.6~$\rm\AA$ (Fe, Ti, Ca, Cr, Ni, and Nd abundances varied to adjust)}\\
 Ti~{\sc ii} & 5188.680 &  1.582 & $-$1.210 & 8.23 & $-$6.66 & $-$7.95 & 91.7& $-$1.52& $-$2.42 & $-249.6$\\
 Ca~{\sc i}  & 5188.844 &  2.932 & $-$0.090 & 8.42 & $-$4.18 & $-$7.31 &124.2& $-$1.10& $-$2.08 & $-262.9$\\
 Ti~{\sc i}  & 5189.621 &  2.239 & $-$1.017 & 8.06 & $-$5.71 & $-$7.79 &  3.7& $-$0.06& $-$0.07 & \\
 Fe~{\sc i}  & 5191.455 &  3.038 & $-$0.656 & 8.00 & $-$5.47 & $-$7.62 &176.4& $-$1.69& $-$3.68 & $-294.5$\\
 Cr~{\sc i}  & 5192.002 &  3.395 & $-$0.390 & 7.91 & $-$5.32 & $-$7.62 & 23.2& $-$0.33& $-$0.48 & \\
 Fe~{\sc i}  & 5192.343 &  2.998 & $-$0.521 & 8.01 & $-$5.47 & $-$7.62 &204.3& $-$1.64& $-$3.92 & $-282.3$\\
 Ni~{\sc i}  & 5192.490 &  3.699 & $-$1.512 & 8.08 & $-$5.67 & $-$7.69 & 26.7& $-$0.41& $-$0.60 & \\
 Nd~{\sc ii} & 5192.614 &  1.136 &  0.290 & 7.92 & $-$5.80 & $-$7.73   &  8.7& $-$0.15& $-$0.22 & \\
 Ti~{\sc i}  & 5192.969 &  0.021 & $-$1.006 & 6.78 & $-$6.32 & $-$7.84 & 84.6& $-$1.78& $-$2.63 & $-247.1$\\
 Cr~{\sc i}  & 5193.480 &  3.422 & $-$0.720 & 6.83 & $-$6.11 & $-$7.84 & 11.9& $-$0.16& $-$0.23 & \\
%\\
\multicolumn{10}{l}{Region (b): 5193.6--5199.0~$\rm\AA$ (Fe, Cr, Mn, Ni, Ti, and Y abundances varied to adjust)}\\
 Ti~{\sc i}  & 5194.034 &  2.103 & $-$0.560 & 7.83 & $-$5.37 & $-$7.62 & 10.2& $-$0.13& $-$0.19 & \\
 Fe~{\sc i}  & 5194.941 &  1.557 & $-$2.090 & 6.29 & $-$6.20 & $-$7.87 &120.7& $-$2.26& $-$3.74 & $-298.9$\\
 Fe~{\sc i}  & 5195.468 &  4.220 & $-$0.002 & 8.31 & $-$6.17 & $-$7.67 &102.9& $-$1.43& $-$2.40 & $-277.0$\\
 Fe~{\sc i}  & 5196.077 &  4.256 & $-$0.451 & 8.33 & $-$5.16 & $-$7.67 & 77.4& $-$1.16& $-$1.80 & $-240.6$\\
 Y~{\sc ii}  & 5196.422 &  1.748 & $-$0.880 & 7.92 & $-$5.93 & $-$7.77 & 22.5& $-$0.37& $-$0.53 & \\
 Cr~{\sc i}  & 5196.482 &  3.449 & $-$0.270 & 7.80 & $-$5.93 & $-$7.82 & 21.1& $-$0.30& $-$0.44 & \\
 Cr~{\sc i}  & 5196.590 &  3.449 & $-$0.360 & 7.83 & $-$5.94 & $-$7.82 & 18.0& $-$0.25& $-$0.37 & \\
 Mn~{\sc i}  & 5196.593 &  3.135 & $-$0.930 & 7.77 & $-$6.09 & $-$7.83 &  9.5& $-$0.13& $-$0.19 & \\
 Ni~{\sc i}  & 5197.157 &  3.898 & $-$1.190 & 8.14 & $-$5.59 & $-$7.81 & 21.7& $-$0.32& $-$0.47 & \\
 Fe~{\sc ii} & 5197.577 &  3.230 & $-$2.100 & 8.48 & $-$6.60 & $-$7.95 & 86.8& $-$1.17& $-$1.93 & $-250.3$\\
 Fe~{\sc i}  & 5197.929 &  4.301 & $-$1.340 & 8.33 & $-$4.77 & $-$7.77 & 31.7& $-$0.47& $-$0.69 & \\
 Fe~{\sc i}  & 5198.711 &  2.223 & $-$2.135 & 8.21 & $-$6.18 & $-$7.83 & 87.5& $-$1.67& $-$2.55 & $-253.3$\\
%\\
\multicolumn{10}{l}{Region (c): 5199.0--5205.0~$\rm\AA$ (Cr, Fe, Y, Nd, and Ti abundances varied to adjust)}\\
 Fe~{\sc i}  & 5199.531 &  3.882 & $-$3.236 & 8.01 & $-$5.47 & $-$7.69 &  1.5& $-$0.03& $-$0.04 & \\
 Fe~{\sc i}  & 5199.695 &  3.546 & $-$3.248 & 7.69 & $-$6.11 & $-$7.80 &  2.9& $-$0.05& $-$0.06 & \\
 Nd~{\sc ii} & 5200.121 &  0.559 & $-$0.490 & 7.92 & $-$5.89 & $-$7.75 & 10.4& $-$0.18& $-$0.27 & \\
 Cr~{\sc i}  & 5200.207 &  3.385 & $-$0.660 & 7.91 & $-$5.34 & $-$7.62 & 14.7& $-$0.20& $-$0.29 & \\
 Y~{\sc ii}  & 5200.406 &  0.992 & $-$0.570 & 7.92 & $-$6.03 & $-$7.80 & 31.5& $-$0.56& $-$0.83 & \\
 Ti~{\sc i}  & 5201.119 &  2.092 & $-$0.749 & 7.83 & $-$5.37 & $-$7.62 &  6.3& $-$0.08& $-$0.12 & \\
 Fe~{\sc i}  & 5202.251 &  4.256 & $-$0.639 & 8.32 & $-$5.88 & $-$7.67 & 66.7& $-$1.02& $-$1.55 & $-229.9$\\
 Fe~{\sc i}  & 5202.335 &  2.176 & $-$1.838 & 8.23 & $-$6.18 & $-$7.83 &103.8& $-$1.89& $-$3.02 & $-279.5$\\
 Cr~{\sc i}  & 5204.506 &  0.941 & $-$0.208 & 7.72 & $-$6.15 & $-$7.86 &164.5& $-$2.22& $-$4.28 & $-307.6$\\
 Fe~{\sc i}  & 5204.582 &  0.087 & $-$4.332 & 3.76 & $-$6.33 & $-$7.89 & 82.3& $-$1.98& $-$2.91 & $-231.0$\\
%\\
\multicolumn{10}{l}{Region (d): 5205.0--5211.0~$\rm\AA$ (Cr, Fe, Ti, and Y abundances varied to adjust)}\\
 Fe~{\sc i}  & 5205.309 &  4.260 & $-$2.673 & 7.99 & $-$5.16 & $-$7.67 &  4.0& $-$0.06& $-$0.08 & \\
 Y~{\sc ii}  & 5205.724 &  1.033 & $-$0.340 & 7.91 & $-$6.03 & $-$7.80 & 43.7& $-$0.82& $-$1.23 & \\
 Cr~{\sc i}  & 5206.038 &  0.941 &  0.019 & 7.72 & $-$6.15 & $-$7.86   &208.6& $-$2.08& $-$4.65 & $-292.5$\\
 Ti~{\sc i}  & 5206.119 &  2.487 &  1.070 & 7.80 & $-$6.12 & $-$7.84   & 51.7& $-$0.85& $-$1.24 & \\
 Cr~{\sc i}  & 5206.203 &  2.708 & $-$1.283 & 8.36 & $-$5.46 & $-$7.57 & 17.4& $-$0.24& $-$0.36 & \\
 Cr~{\sc i}  & 5206.561 &  3.435 & $-$1.251 & 6.98 & $-$5.89 & $-$7.82 &  4.5& $-$0.07& $-$0.09 & \\
 Fe~{\sc i}  & 5206.594 &  4.294 & $-$2.908 & 7.90 & $-$4.52 & $-$7.59 &  2.2& $-$0.04& $-$0.05 & \\
 Fe~{\sc i}  & 5206.801 &  4.283 & $-$2.530 & 8.31 & $-$4.32 & $-$7.76 &  5.3& $-$0.07& $-$0.10 & \\
 Fe~{\sc i}  & 5207.939 &  3.635 & $-$2.450 & 8.30 & $-$6.05 & $-$7.84 & 20.7& $-$0.31& $-$0.46 & \\
 Cr~{\sc i}  & 5208.112 &  2.709 & $-$1.772 & 8.36 & $-$5.46 & $-$7.57 &  6.4& $-$0.09& $-$0.12 & \\
 Cr~{\sc i}  & 5208.419 &  0.941 &  0.158 & 7.72 & $-$6.15 & $-$7.86   &234.0& $-$1.99& $-$4.81 & $-279.7$\\
 Fe~{\sc i}  & 5208.593 &  3.241 & $-$0.980 & 7.88 & $-$5.51 & $-$7.62 &114.7& $-$1.64& $-$2.84 & $-291.2$\\
 Fe~{\sc i}  & 5209.884 &  3.237 & $-$3.460 & 7.31 & $-$6.29 & $-$7.87 &  5.9& $-$0.09& $-$0.12 & \\
 Ti~{\sc i}  & 5210.386 &  0.048 & $-$0.884 & 6.77 & $-$6.31 & $-$7.84 & 73.1& $-$1.52& $-$2.21 & $-227.8$\\
 Cr~{\sc ii} & 5210.865 &  3.758 & $-$2.945 & 8.42 & $-$6.46 & $-$7.89 &  5.0& $-$0.07& $-$0.10 & \\
\hline
\end{tabular}
\end{center}
%\scriptsize
\tiny
Shown here are the atomic data of the lines of appreciable strengths
(see also Figure 2), which were taken from Kurucz \& Bell's (1995) compilation 
(with adjustments for 4 lines; {\it cf.} Section 3.2). Followed by first 
four self-explanatory columns, damping parameters are presented in columns 5--7:
Gammar is the radiation damping width (s$^{-1}$), $\log\gamma_{\rm rad}$.
Gammas is the Stark damping width (s$^{-1}$) per electron density (cm$^{-3}$) 
at $10^{4}$ K, $\log(\gamma_{\rm e}/N_{\rm e})$.
Gammaw is the van der Waals damping width (s$^{-1}$) per hydrogen density 
(cm$^{-3}$) at $10^{4}$ K, $\log(\gamma_{\rm w}/N_{\rm H})$. 
In case where these damping parameters are not given in Kurucz and Bell
(1995), they were computed as default values in Kurucz's WIDTH program
(Leushin and Topil'skaya, 1987). Columns 8--10 give the strength-related
quantities relevant to the disk-center spectrum ({\it cf.} Section 3.3): 
the equivalent width inversely computed from the best-fit abundance 
solution ({\it cf.} Figure 2) under the assumption of isolated single line, 
the mean depth of line formation, and the line-center formation depth. 
Presented in column 11 is the theoretically estimated convective 
blue-shift at $\mu=1$ (only for the representative 18 lines; {\it cf.} Section 5.2).
\end{table}

%% Figure 1
\begin{figure} 
\centerline{\includegraphics[width=0.7\textwidth]{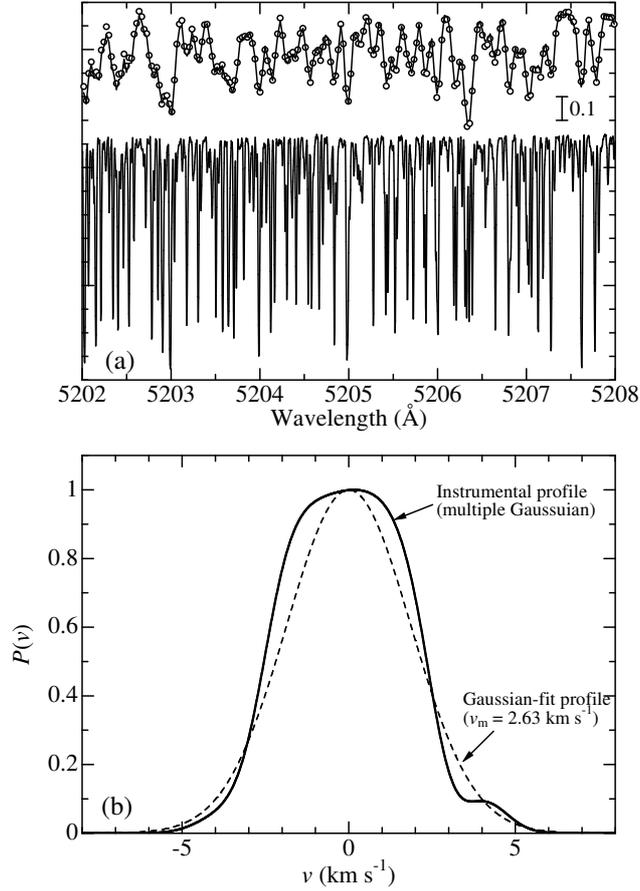}}
\caption{Determination of the instrumental profile ($P_{\lambda}$)
by comparing the observed lamp+I$_{2}$ spectrum ($D_{\lambda}^{\rm obs}$)
with the simulated spectrum 
($D_{\lambda}^{\rm sim} \equiv T_{\lambda} \otimes P_{\lambda}$, where 
$T_{\lambda}$ is the transmission function of I$_{2}$ vapor.
(a) Upper spectra are the observed $D_{\lambda}^{\rm obs}$ (open circles)
with the best-fit $D_{\lambda}^{\rm sim}$ (solid line) overplotted,
while the lower spectrum (solid line) is $T_{\lambda}$.
(b) Thick solid line is the finally obtained $P(v)$
(the sum of nine off-center Gaussian functions; {\it cf.} Section 2),
while the dashed line is the special case derived by assuming 
that $P$ is the Gaussian form. The abscissa is in the velocity scale, 
and both profiles are normalized at the maximum peak to be unity.
}%\label{}
\end{figure}

%% Figure 2
\begin{figure} 
\centerline{\includegraphics[width=0.8\textwidth]{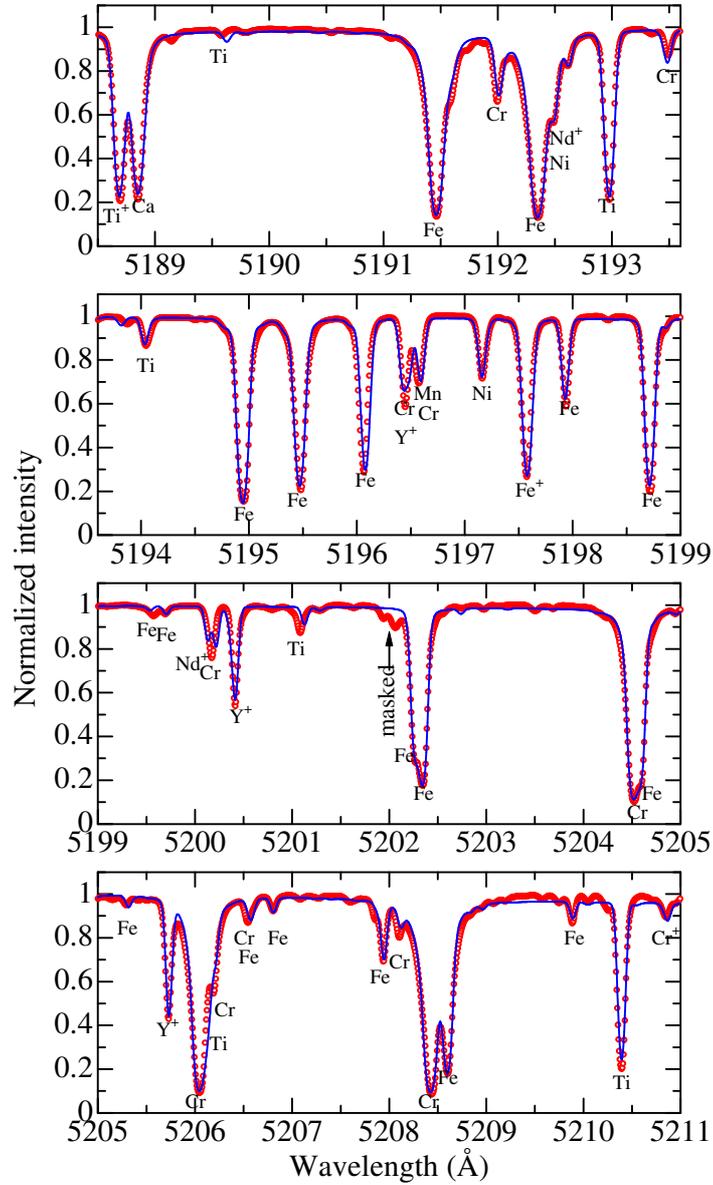}}
\caption{
Demonstration of how the solar intensity spectrum simulated at $\mu = 1$ 
for each wavelength region (a--d) can be fitted with the observed solar 
disk-center spectrum (Neckel, 1994, 1999) by adjusting the abundances 
and the macrobroadening parameter. Observed and theoretical spectrum 
are expressed by (red) open circles and (blue) solid lines, 
respectively. Identifications are given for the line features of 
appreciable strengths.
}%\label{}
\end{figure}

%% Figure 3
\begin{figure} 
\centerline{\includegraphics[width=0.8\textwidth]{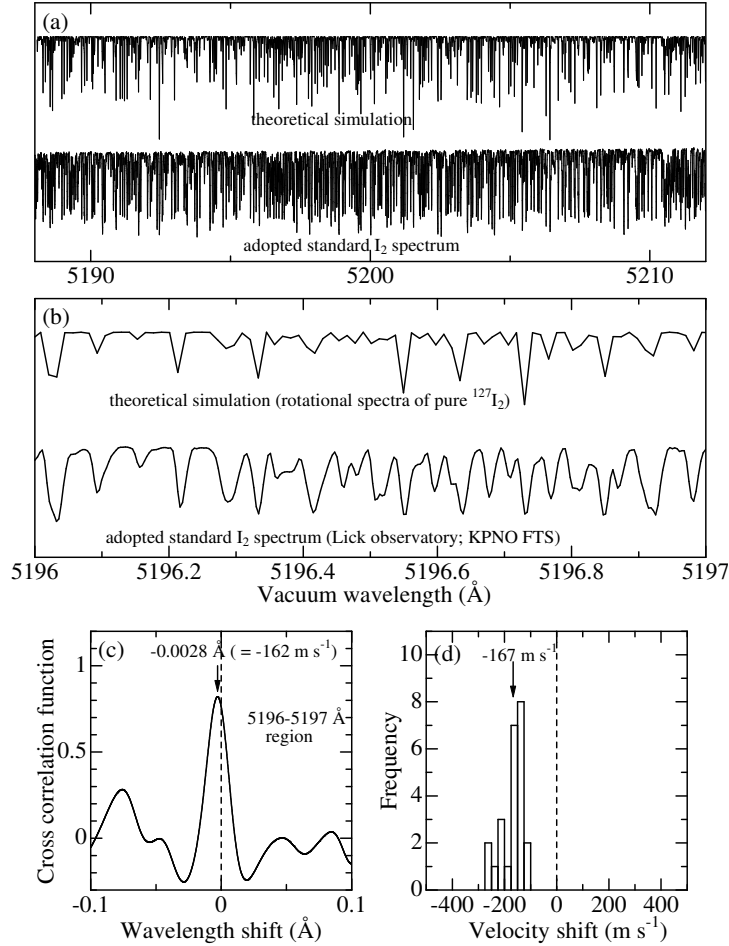}}
\caption{
Comparison of the theoretical pure $^{127}$I$_{2}$ absorption spectrum
simulated by Kn\"{o}ckel {\it et al.}'s (2004) ``IodineSpec'' program with 
the observed I$_{2}$ absorption spectrum (Lick I$_{2}$ cell spectrum 
recorded with Kitt Peak FTS) adopted as the reference in this study: 
(a) wide view for the 5188--5212~$\rm\AA$ region and (b) magnified view 
for the 5196--5197~$\rm\AA$ region (note that vacuum wavelength scale
is used in these two figures).
(c) Result of cross-correlation analysis for the 5196--5197~$\rm\AA$ 
region shown in panel (b), where the abscissa corresponds to the 
wavelength correction to be applied to the observed spectrum
to enable a match between the two. 
(d) Histogram showing the distribution of the corrections
(expressed in the velocity scale) derived by cross-correlation analyses 
for each of the twenty-four 1~$\rm\AA$ bins in the 5188--5212~$\rm\AA$ 
region.
}%\label{}
\end{figure}

%% Figure 4
\begin{figure} 
\centerline{\includegraphics[width=0.8\textwidth]{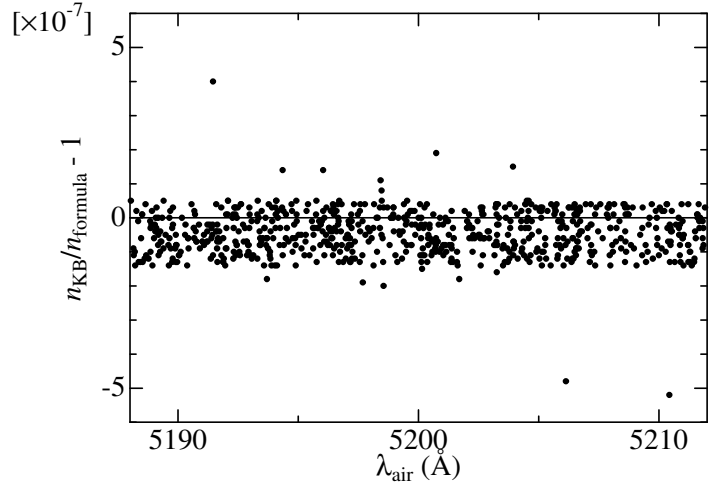}}
\caption{
Wavelength-dependence of the relative difference between $n_{\rm formula}$ 
and $n_{\rm KB}$ (two kinds of refractive indices) for all the 
spectral lines at $5188\rm\AA \le \lambda_{\rm air} \le 5212\rm\AA$
included in Kurucz and Bell's (1995) compilation, 
where $n_{\rm formula}$ is the refractive index of air based on the standard 
formula given in Cox (2000) as given in Equation (7) while $n_{\rm KB}$ is 
estimated from Kurucz and Bell's (1995) atomic data by Equation (8).
}%\label{}
\end{figure}

%% Figure 5
\begin{figure} 
\centerline{\includegraphics[width=0.8\textwidth]{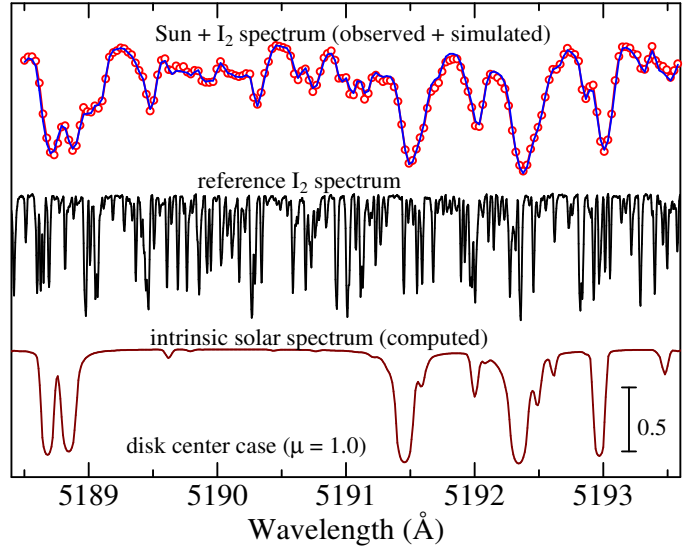}}
\caption{
Graphical example [for the disk-center case at region (a)] of 
how the absolute wavelength shift can be determined by searching 
for the best-fit parameters such those accomplishing the best-fit 
(simulated) $I_{\lambda, {\rm th}}$ [$\equiv G_{\lambda} \otimes 
(T_{\lambda} \cdot I_{\lambda, {\rm intr}})$; {\it cf.} Equation (2)] 
with the observed $I_{\lambda, {\rm obs}}$ 
(observed solar intensity spectrum recorded through the I$_{2}$ 
absorption cell. Top: comparison of $I_{\lambda, {\rm obs}}$
(red open circles) and $I_{\lambda, {\rm th}}$ (blue solid line). 
Middle: $T_{\lambda}$ (transmission spectrum of I$_{2}$ vapor
cell). Bottom: $I_{\lambda, {\rm intr}}$ (computed intrinsic 
solar intensity spectrum).
}%\label{}
\end{figure}

%% Figure 6
\begin{figure} 
\centerline{\includegraphics[width=1.\textwidth]{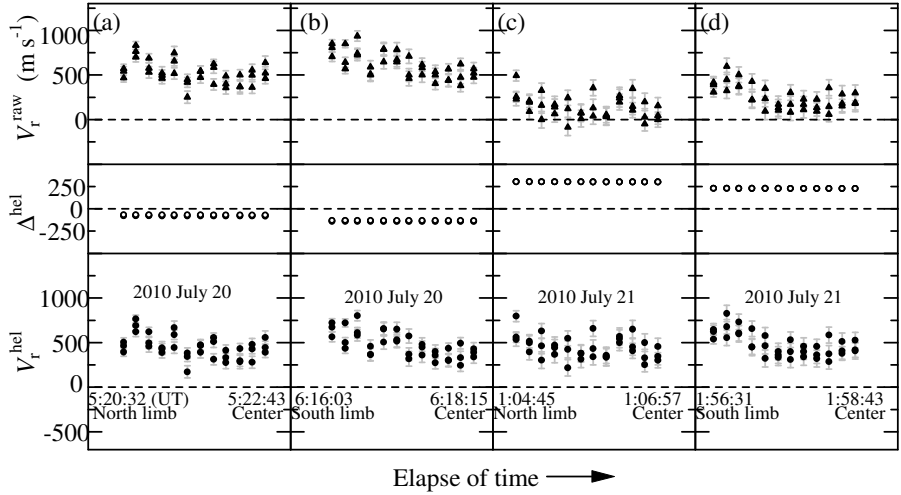}}
\caption{
Runs of derived raw radial velocities ($V_{\rm r}^{\rm raw}$),
heliocentric corrections ($\Delta^{\rm hel}$), and heliocentric
radial velocities ($V_{\rm r}^{\rm hel}$) with time, for four kinds of
scanning observations along the solar meridian line relevant for this study,
each of which were carried out from the limb to the disk center 
over $\approx$~2 minutes. The results for the 2010 July 20 observations 
(east--west aligned slit) are shown in panels (a) ($\theta = 0^{\circ}$; 
northern meridian) and (b) ($\theta = 180^{\circ}$; 
southern meridian). Similarly, those for the 2010 July 21
observations (north--south aligned slit) are presented in panels (c) and (d).
In each panel, the horizontal dashed line denotes the zero-level.
}%\label{}
\end{figure}

%% Figure 7
\begin{figure} 
\centerline{\includegraphics[width=0.8\textwidth]{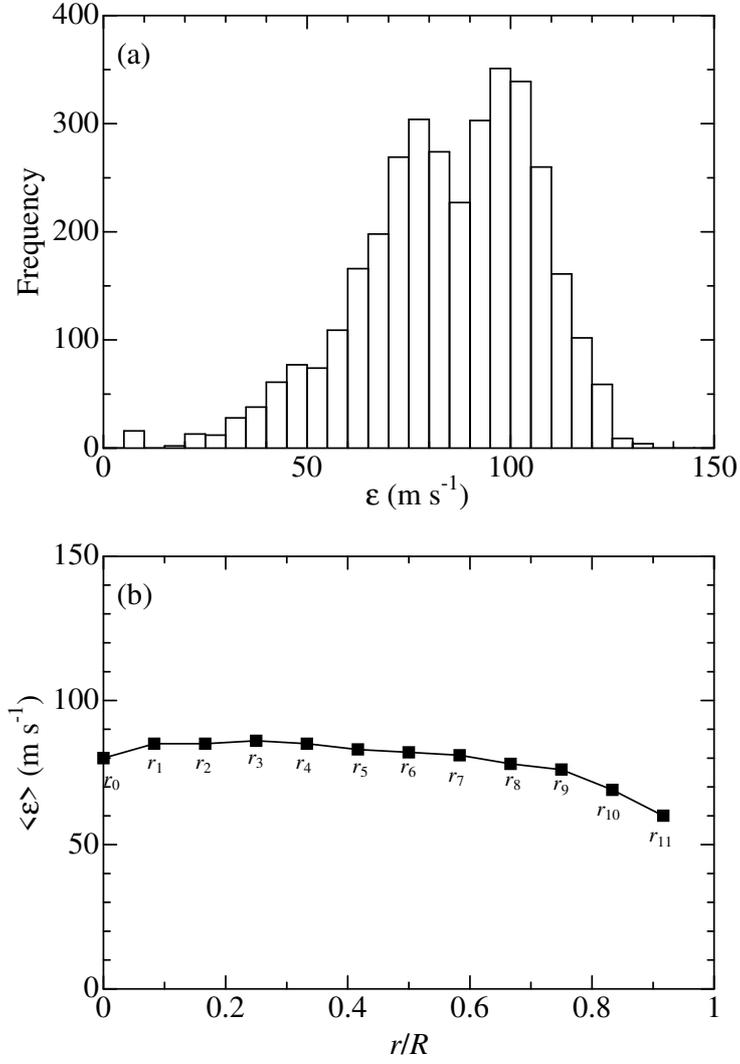}}
\caption{
(a) Distribution histogram for the probable error ($\epsilon$) in 
$V_{\rm r}$ (where results all over the disk are combined).
(b) Mean probable errors ($\langle \epsilon \rangle$; averaged 
at the same radius bin) plotted against the radius.
}%\label{}
\end{figure}

%% Figure 8
\begin{figure} 
\centerline{\includegraphics[width=0.9\textwidth]{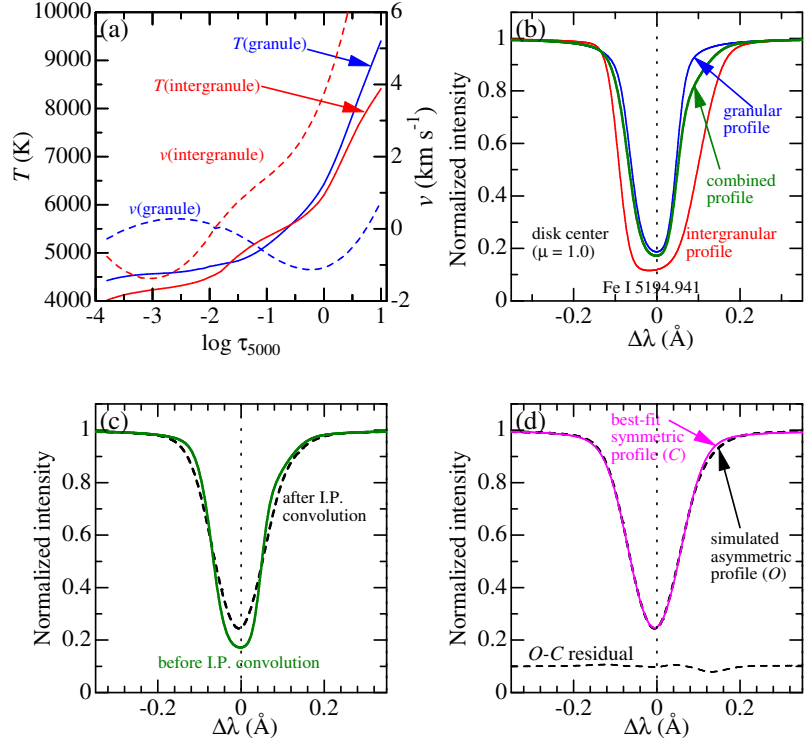}}
\caption{
(a) $T(\tau_{5000})$ (solid line) and $v(\tau_{5000})$ (dashed line;
positive for the downflow) structures of two-component solar
photospheric model (blue and red lines corresponding to granular
and intergranular region, respectively) derived by Borrero and 
Bellot Rubio (2002), which we adopted for the line-profile simulation.
(b) Illustration of how the realistic asymmetric profile (green) is 
produced as a combination of granular (blue) and intergranular (red) 
profiles (being weighted according to the filling factor and the 
continuum intensity), for the case of Fe~{\sc i} 5194.941 line 
at the disk-center.
(c) Example of how the simulated profile (green solid line) 
in (b) is broadened to result in a realistic profile (black dashed 
line) corresponding to our observation by convolving 
the instrumental profile ({\it cf.} thick solid line in Figure 1b).
(d) Demonstration of how the asymmetric profile simulated in (c) 
(black-dashed line) can be fitted with the classically-modeled 
symmetric profile (pink solid line) by adjusting the profile 
parameters (Fe abundance, line broadening parameter $v_{\rm M}$, 
and the line shift). The $O-C$ residual is also shown
at the bottom of the figure (with an offset of +0.1).
}%\label{}
\end{figure}

%% Figure 9
\begin{figure} 
\centerline{\includegraphics[width=0.8\textwidth]{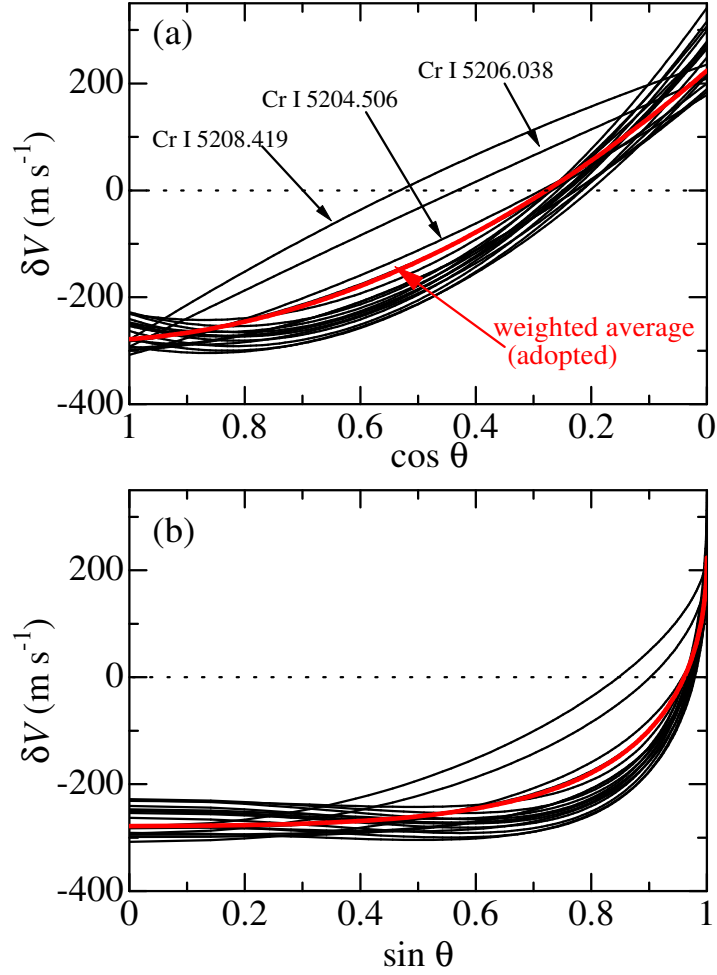}}
\caption{
Angle/position-dependent convective line shifts ($\delta V$) for 
18 representative lines of appreciable strengths, which were simulated 
based on Borrero and Bellot Rubio's (2002) two-component model 
for the disk center ($\delta V_{1}$) and combined with 
Balthasar's (1984) empirical $\delta V_{\mu} - \delta V_{1}$ relation 
({\it cf.} Section 5.3 and Appendix).
(a) $\delta V$ vs. $\cos \theta (\equiv \mu)$, 
(b) $\delta V$ vs. $\sin \theta (\equiv y/R)$.
The finally adopted relation (averaged over 18 lines) is depicted 
by the thick red line.
}%\label{}
\end{figure}

%% Figure 10
\begin{figure} 
\centerline{\includegraphics[width=0.9\textwidth]{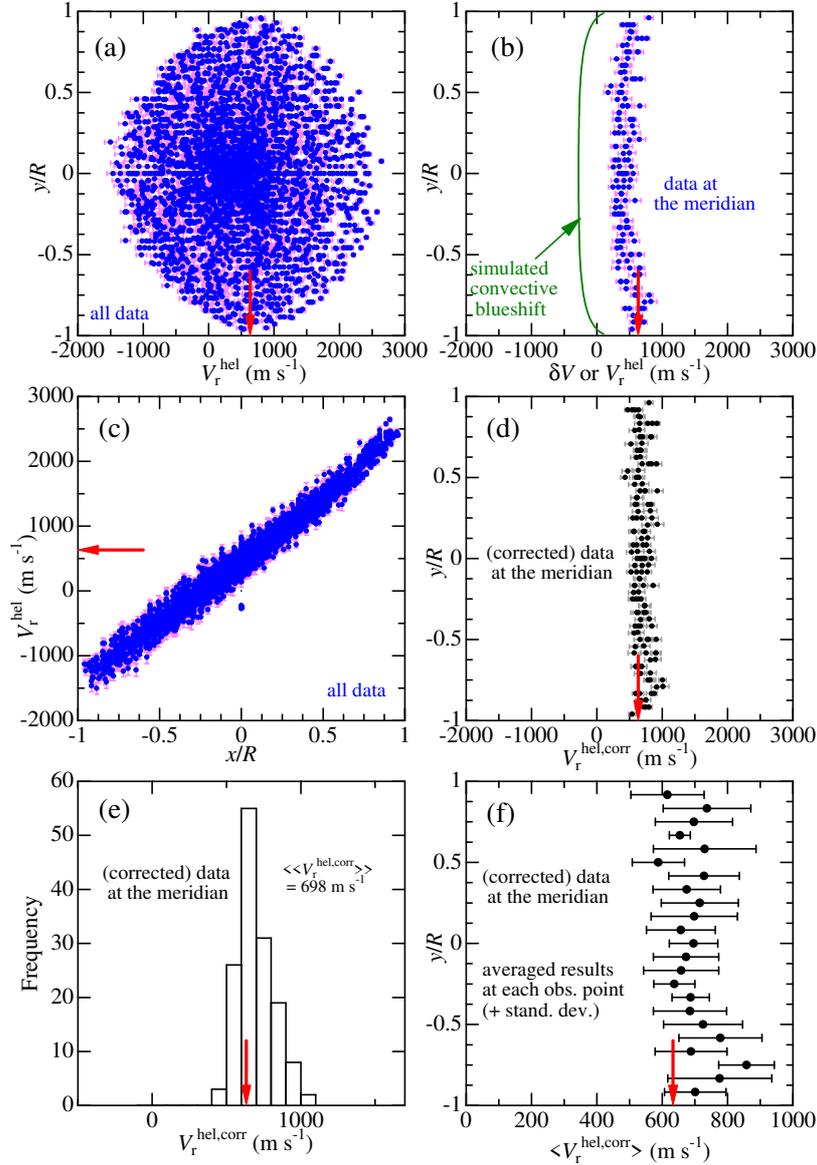}}
\caption{Behaviors of the $V_{\rm r}^{\rm hel}$ (absolute heliocentric
Doppler shift in m~s$^{-1}$) at various positions of the solar disk
($x$ coordinate: E--W direction increasing toward west, 
$y$ coordinate: S--N direction increasing toward north); 
expressed in unit of the solar radius $R$).
(a) $V_{\rm r}^{\rm hel}$ vs. $y$ for all measurements over the entire disk.
(c) $x$ vs. $V_{\rm r}^{\rm hel}$ for all measurements over the entire disk.
(b) $V_{\rm r}^{\rm hel}$ vs. $y$ only for the meridian measurements ($x = 0$).
The positional-dependence of the simulated convective shift 
[$\langle \delta V_{\mu} \rangle$; {\it cf.} Equation (17)], 
which is a function of $y$ through the relation 
$\mu = \sqrt{1-(x^{2} + y^{2})/R^{2}} = \sqrt{1- (y/R)^{2}}$ is 
also depicted by the solid line.
(d) $V_{\rm r}^{\rm hel,corr} 
(\equiv V_{\rm r}^{\rm hel} - \langle \delta V_{\mu} \rangle)$ 
vs. $y$ only for the meridian measurements ($x = 0$).
(e) Histogram showing the distribution of $V_{\rm r}^{\rm hel,corr}$.
(f) $\langle V_{\rm r}^{\rm hel} \rangle$ vs. $y$, where
$\langle V_{\rm r}^{\rm hel} \rangle$ is the average of data 
at almost the same meridian points (bars indicate the standard deviation).
In panels (a)--(d), probable errors involved for each of the $V_{\rm r}$ 
measurements are indicated by thin ticks. The expected solar gravitational 
redshift at the earth (633~m~s$^{-1}$) is shown by the (red) arrow 
in all six panels.
}%\label{}
\end{figure}

%% Figure 11
\begin{figure} 
\centerline{\includegraphics[width=0.9\textwidth]{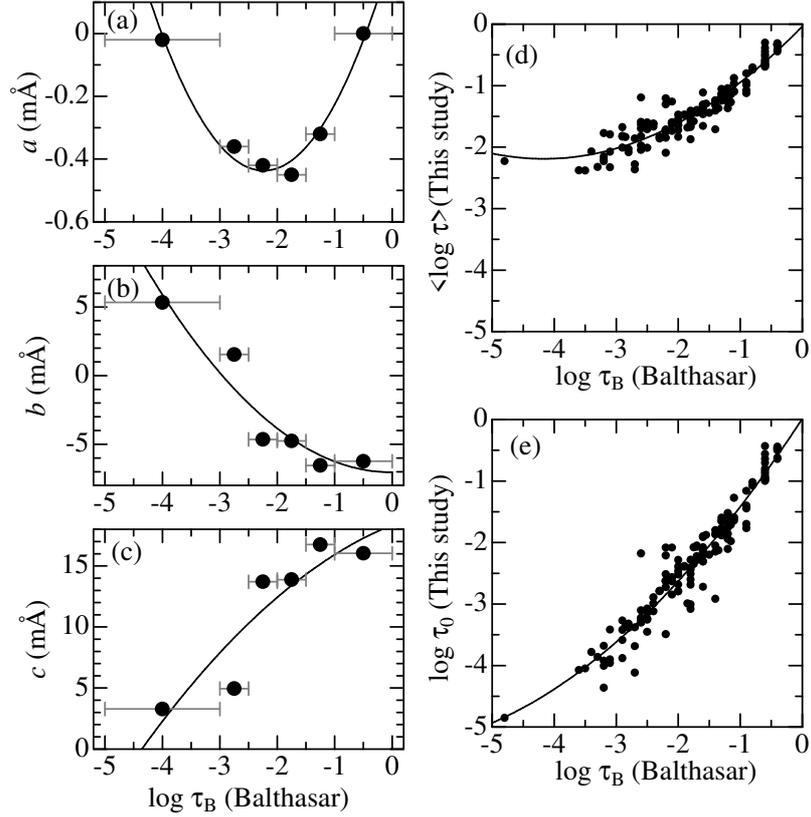}}
\caption{
Panels (a)--(c): Graphical representation of the depth-dependence of 
the coefficients ($a$, $b$, $c$) used for the polynomial fits
of the limb effect, $\Delta \lambda = a + b (1- \mu) + c (1 - \mu)^2$,
which were empirically determined by Balthasar (1984; {\it cf.} Table IV therein). 
The quadratic dependence of these coefficients in terms of $\log \tau_{\rm B}$,
which we derived from the least-squares fit are also depicted.
Panels (d) and (e) show the relationships between the line-core-forming 
depth ($\log \tau_{\rm B}$) adopted by Balthasar (1984) and the two kinds 
of line-formation depths ($\langle \log \tau \rangle$ and $\log \tau_{0}$;
{\it i.e.}, mean depth of line formation and the line-center-forming depth) 
derived in this study. The least-squares fitted quadratic relations
are also drawn.
}%\label{}
\end{figure}

%% Table
%
% \begin{table}
% \caption{}%\label{tbl:?}
% \begin{tabular}{}     
% \hline
% \multicolumn{2}{c}{<>}
% <data>
% \hline
% \end{tabular}
% \end{table}

%%%%%%%%%%%%%%%%%%%%%%%%%%%%%%%%%%%%%%%%%%%%%%%%%%%%%%%%%%%%%%%%%%%%%%%%%%%
%% Appendix
%
% \appendix   

%%%%%%%%%%%%%%%%%%%%%%%%%%%%%%%%%%%%%%%%%%%%%%%%%%%%%%%%%%%%%%%%%%%%%%%%%%%
%% Acknowledgements
%
% \begin{acks}
%
% \end{acks}

%%% %%%%%%%%%%%%%%%%%%%%%%%%%%%%%%%%%%%%%%%%%%%%%%%%%%%%%%%%%%%
%% Bibliography
%
% Using BibTeX
%
% \bibliographystyle{spr-mp-sola}
% %\bibliographystyle{spr-mp-sola-cnd} %% Alternative style: no title, no concluding page
% \bibliography{<bib file>}  
%
% Without BibTeX 
% \begin{thebibliography}{}
% \bibitem[\protect\citeauthoryear{Author}{Year}]{key}
%   <bibliographical entry>
%
% \bibitem[\protect\citeauthoryear{}{}]{}
%   
%  
% \end{thebibliography}

\end{article} 
\end{document}